\documentclass[twocolumn,showkeys,preprintnumbers,amsmath,amssymb,nofootinbib]{revtex4}

\IfFileExists{srcltx.sty}{\usepackage[active]{srcltx}} 

\usepackage{graphicx}
\usepackage{dcolumn}
\usepackage{bm}

\begin{document}

\preprint{\small
Int.J.Mod.Phys. D16(12a) (2007) 1879-1925
}

\title{Space-based research in fundamental physics and quantum technologies}

\author{Slava G. Turyshev, Ulf E. Israelsson, Michael Shao, Nan Yu}
\affiliation{%
Jet Propulsion Laboratory, California Institute of Technology,\\
4800 Oak Grove Drive, Pasadena, CA 91109-0899, USA
}%

\author{Alexander Kusenko, Edward L. Wright}
\affiliation{%
Department of Physics and Astronomy, University of California, Los Angeles, CA 90095-1547, USA
}%

\author{C.W. Francis Everitt, Mark A. Kasevich, John A. Lipa, John C. Mester}
\affiliation{%
Hansen Experimental Physics Laboratory, Department of Physics,
Stanford University, Stanford, CA 94305-4085, USA
}%

\author{Robert D. Reasenberg, Ronald L. Walsworth}
\affiliation{%
Harvard-Smithsonian Center for Astrophysics, 60 Garden Street, Cambridge, MA 02138, USA
}%

\author{Neil Ashby}
\affiliation{%
Department of Physics, University of Colorado, Boulder, CO 80309-0390, USA
}%

\author{Harvey Gould}
\affiliation{%
Lawrence Berkeley National Laboratory, One Cyclotron Road, Berkeley, CA 94720, USA
}%

\author{Ho Jung Paik}
\affiliation{%
Department of Physics, University of Maryland, College Park, MD 20742-4111, USA
}%

\date{\today}

\begin{abstract}
Space offers unique experimental conditions and a wide range of opportunities to explore the foundations of modern physics with an accuracy far beyond that of ground-based experiments. Space-based experiments today can uniquely address important questions related to the fundamental laws of Nature. In particular, high-accuracy physics experiments in space can test relativistic gravity and probe the physics beyond the Standard Model; they can perform direct detection of gravitational waves and are naturally suited for precision investigations in cosmology and astroparticle physics. In addition, atomic physics has recently shown substantial progress in the development of optical clocks and atom interferometers. If placed in space, these instruments could turn into powerful high-resolution quantum sensors greatly benefiting fundamental physics. 

We discuss the current status of space-based research in fundamental physics, its discovery potential, and its importance for modern science.   We offer a set of recommendations to be considered by the upcoming National Academy of Sciences' Decadal Survey in Astronomy and Astrophysics. In our opinion, the Decadal Survey should include space-based research in fundamental physics as one of its focus areas. We recommend establishing an Astronomy and Astrophysics Advisory Committee's interagency ``Fundamental Physics Task Force''  to assess the status of both ground- and space-based efforts in the field, to identify the most important objectives, and to suggest the best ways to organize the work of several federal agencies involved.  We also recommend establishing a new NASA-led interagency program in fundamental physics that will consolidate new technologies, prepare key instruments for future space missions, and build a strong scientific and engineering community. Our goal is to expand NASA's science objectives in space by including ``laboratory research in fundamental physics'' as an element in agency's ongoing space research efforts. 
\end{abstract}

\maketitle

\section{\label{sec:intro}Introduction}

Today physics stands at the threshold of major discoveries.  Growing
observational evidence points to the need for new physics. Efforts to discover new fundamental symmetries, investigations of the limits of established symmetries, tests of the general theory of relativity, searches for gravitational waves, and attempts to understand the nature of dark matter were among the topics that had been the focus of the scientific research at the end of the last century.  These efforts have further intensified with the discovery of dark energy made in the late 1990s, which triggered many new activities aimed at answering important questions related to the
most fundamental laws of Nature \cite{Wilczek-2007}.

The 2003 Report {\it ``Connecting Quarks with the Cosmos: Eleven Science
Questions for the New Century''},\footnote{{\it ``Connecting Quarks with the
Cosmos: Eleven Science Questions for the New Century,''} Board on Physics and
Astronomy (The National Academies Press, 2003).  In particular, the report
identified the following eleven questions that are shaping the modern research
in astronomy, astrophysics, and cosmology:  i) What is the dark matter? ii)
What is the nature of the dark energy?  iii) How did the universe begin?  iv)
Was Einstein right about gravity?  v) How have neutrinos shaped the universe?
vi) What are nature's most energetic particles? vii) Are protons unstable?
viii) What are the new states of matter?  ix) Are there more space-time
dimensions?  x) How were elements from iron to uranium made?  xi) Is a new
theory of matter and light needed?} issued by the National Academy of Science's
Board of Physics and Astronomy, identified the most critical research areas
that require support to resolve the profound challenges facing physics and
astronomy today.  The report became a blueprint for multi-agency efforts at
the National Aeronautics and Space Administration (NASA), the National Science
Foundation (NSF), and the U.S. Department of Energy (DOE), aiming at overcoming
the evident challenges in our understanding of matter, space, time, and the
universe.  
Although the report provided a list of strategic recommendations allowing NASA, NSF, and DOE to select various fundamental physics projects among their top priorities, more work is needed.  In addition, some of the science opportunities, including those offered by space-based laboratory research in fundamental physics\footnote{Reprioritization of
space efforts initiated by NASA in 2004, led to the termination of a successful {\it ``Microgravity and Fundamental Physics''} program managed by the former Office of Biological and Physics Research.  As a result, no program for space-based laboratory research in fundamental physics currently exists.}, were overlooked.

Historically, the nature of matter on Earth and the laws governing it were discovered in laboratories on Earth. To understand the nature of matter in the universe and the laws governing it is reasonable that we move our laboratories outside the Earth. There are two approaches to physics research in space: one can detect and study signals from remote astrophysical objects (the ``observatory'' mode) or one can perform carefully designed experiments in space (the ``laboratory'' mode).  The two methods are complementary and the latter, which is the focus of this paper, has the advantage of utilizing the well-understood and controlled environments of a space-based laboratory. Existing technologies allow one to take advantage of the unique environments found only in space, including variable gravity potentials, large distances, high velocity and low acceleration regimes, availability of pure geodetic trajectories, microgravity and thermally-stable environments (see Appendix~\ref{sec:2} for details).  With recent advances in several applied physics disciplines new instruments and technologies have become available. These include highly accurate atomic clocks, optical frequency combs, atom interferometers, drag-free technologies, low-thrust micro-propulsion techniques, optical transponders, long-baseline optical interferometers, etc. \cite{Phillips-2007}. Some of these instruments are already space-qualified, thereby enabling a number of high-precision investigations in laboratory fundamental physics in space. As a result, space-based experiments are capable of reaching very high accuracies in testing the foundations of modern physics.  Furthermore, because experimental physics research complements the observational disciplines of astronomy and astrophysics, it is possible that independent confirmation by space-based fundamental physics experiments may be required to fully explain any future observations of, for example, ``detection'' of dark matter particles
or identification of the source for dark energy.

As was demonstrated at the two recent international ``Quantum to Cosmos'' workshops,\footnote{{\it ``From Quantum to Cosmos: Fundamental Physics Research in Space''}, Airlie Center, Warrenton, VI, USA, May 21-24, 2006, {\tt
http://physics.jpl.nasa.gov/quantum-to-cosmos}; \\
{\it ``From Quantum to Cosmos -- II: Space-Based Research in Fundamental Physics \& Quantum Technologies''}, Bremen, Germany, June 10-13, 2007, {\tt
http://www.zarm.uni-bremen.de/Q2C2/}
} there is a growing community of researchers worldwide interested in performing carefully thought out laboratory physics experiments to address some of the modern challenges that physics faces today, by utilizing the benefits of a space environment. The recent report of the Committee on Atomic, Molecular, and Optical Sciences (AMO2010)\footnote{{\it ``Controlling the Quantum World,''} Committee on Atomic, Molecular, and Optical Sciences (AMO2010), Board on Physics and Astronomy (The National Academies of Press, July 2006). Electronic version of the report is available from NAP website at {\tt http://www.nap.edu/catalog.php?record\_id=11705}} emphasized the significant discovery potential of future space-based experiments using new technologies and laboratory techniques, especially in their ability to probe the fundamental laws of Nature at the highest levels of accuracy.  The 2006 Report from the Dark Energy Task Force also endorsed the important role of gravitational experiments as an effective means to discover new physics that might also be at play on cosmological scales and that might be responsible for the small observed acceleration of the cosmological expansion of the universe.\footnote{{\it ``Report from the Dark Energy Task Force,''} June 6, 2006, at [arXiv:astro-ph/0609591]; electronic version of the Report is available at {\tt http://www.nsf.gov/mps/ast/aaac/dark\_energy\_
task\_force/report/detf\_final\_report.pdf}} 

The 2005 Position Paper by the European Physical Society (EPS)\footnote{{\it ``The Need for Space Flight Opportunities in Fundamental Physics,''} a Position Paper of the European Physical Society (EPS), published on the occasion of the centenary of Albert Einstein's annus mirabilis (2005) and available from EPS' website: {\tt
http://www.eps.org/papers\_position/paper\_index.html}} highlighted strong
discovery potential of space-based experiments in fundamental physics and
argued for space flight opportunities specifically dedicated to this area of
research.  The EPS' recommendations further supported the efforts of the European Space Agency's (ESA) Fundamental Physics Advisory Group (FPAG)\footnote{\label{fpag}The webpage of the ESA's Fundamental Physics Advisory Group (FPAG): {\tt
http://sci.esa.int/science-e/www/object/ index.cfm?fobjectid=33212}} -- an
influential group of European scientists that advises ESA on scientific
direction in fundamental physics research in space.  As a result, ESA's Cosmic Vision 2015-2025 process\footnote{\label{esa-cv}For details on the ESA's Cosmic Vision 2015-2025 process and the recent Call for Mission Proposals, visit {\tt http://sci.esa.int/ science-e/www/object/index.cfm?fobjectid=40794}} marks a breakthrough for fundamental physics: for the first time, a major space agency has given full emphasis in its forward planning to missions dedicated to exploring and
advancing the limits of our understanding of many fundamental physics issues,
including gravitation, unified theories, and quantum theory. NASA would benefit from a similar bold and visionary approach. 

This paper is organized as follows.  In Section~\ref{sec:3} we discuss the status of space-based research in fundamental physics, present examples of experiments that could provide significant advances in the field in the near future, and emphasize scientific and societal benefits of this space science discipline. Each subsection discusses the significance of physics to be addressed, emphasizes the role of space for a particular kind of experiment, and presents a list of potential missions. 
  In Section~\ref{sec:4} we argue for a coordinated multi-agency support for space-based research in fundamental physics and present a set of policy recommendations which, if adopted, would re-energize the entire field of research in fundamental physics.  In Appendix~\ref{sec:2} we present the benefits of space-based deployment for precision physics experiments.  In Appendix~\ref{sec:2app} we discuss the history of fundamental physics research at NASA and its current programmatic status.

\section{\label{sec:3}Fundamental Physics in Space: Great Potential for Discovery}

The fundamental physical laws of Nature are currently described by the Standard Model and Einstein's general theory of relativity.  The Standard Model specifies the families of fermions (leptons and quarks) and their interactions by vector fields which transmit the strong, electromagnetic, and weak forces. General relativity is a tensor field theory of gravity with universal coupling to the particles and fields of the Standard Model.

Despite the beauty and simplicity of general relativity and the success of the Standard Model, our present understanding of the fundamental laws of physics has several shortcomings. Although recent  progress in string theory \cite{Witten-2001} is very encouraging,  the search for a realistic theory of quantum gravity remains a challenge. This continued inability to merge gravity with quantum mechanics indicates that the pure tensor gravity of general relativity needs modification or augmentation. It is now believed that new physics is needed to resolve this issue. Recent work in scalar-tensor extensions of gravity, brane-world gravitational models, and also efforts to modify gravity on large scales motivate new searches for experimental signatures of very small deviations from general relativity on various scales, including on the spacecraft-accessible distances in the solar system.

In addition, the Higgs boson, a particle predicted by the Standard Model, has yet to be discovered.  It is widely expected that the Large Hadron Collider (LHC) at CERN  will be able to probe the nature of electroweak symmetry breaking and verify the prediction of the Higgs boson in the near future.  In addition to this long-anticipated discovery, one hopes to find new physics beyond the Standard Model at the LHC.  The new physics could explain the hierarchy of scales and resolve the naturalness problems associated with the Standard Model.  Physics beyond the Standard Model is required to explain dark matter and the matter-antimatter asymmetry of the universe.  Furthermore,  the Standard Model does not offer an explanation to the observed spectrum of fermion masses and their mixing angles.    The exact conservation of the Charge-conjugation and Parity (CP) symmetries in strong interactions appears mysterious in the Standard Model, because it requires the exact cancellation of two seemingly unrelated contributions to the measurable quantity $\theta$, which {\em a priori} can take any value between 0 and $2\pi$.  New physics is expected to shed light on this mystery as well.

Theoretical models of the kinds of new physics that can solve the problems above typically involve new interactions, some of which could manifest themselves as violations of the equivalence principle, variation of fundamental constants, modification of the inverse square law of gravity at short distances, Lorenz symmetry breaking, as well as large-scale gravitational phenomena.   Each of these manifestations offers an opportunity for space-based experimentation and, hopefully, a major discovery.

Our objective is to emphasize the uniqueness and advantages of space as an experimental site when addressing the challenges above and, thereby, to demonstrate that space-based laboratory research in fundamental physics is a unique area of space science that offers science investigations of the highest quality.

In the subsections below we discuss the current status of space-based research in fundamental physics, including gravitational experiments, the search for new physics beyond the Standard Model, efforts at direct detection of gravitational waves and their use as a probe of physics in the strong gravitational limit, and also cosmology, astroparticle, and atomic physics.

\subsection{\label{sec:2.1}Search for a new theory of gravity and cosmology with experiments in space}

The recent remarkable progress in observational cosmology has subjected general theory of relativity to increased scrutiny by suggesting a non-Einsteinian model of the universe's evolution. From a theoretical standpoint, the challenge is even stronger -- if gravity is to be quantized, general relativity will have to be modified. Furthermore, recent advances in the scalar-tensor extensions of gravity \cite{Damour-Nordtvedt-1993,Damour-Polyakov-1994,Damour-Piazza-Veneziano-2002} have motivated searches for very small deviations from Einstein's theory, at the level of three to five orders of magnitude below the level currently tested by experiment \cite{Turyshev-etal-2004,Hellings-2007}. For many of the modern gravitational experiments, space is an essential laboratory that, in combination with modern technologies, offers unique conditions that are much purer than those achievable in the best ground-based laboratories \cite{DeBra-1997,Laemmerzahl-2004}.

Below we discuss a number of laboratory experiments that benefit from the space deployment. 

\subsubsection{Test of Einstein's Equivalence Principle}

The Einstein's Equivalence Principle (EP)  \cite{Damour-Nordtvedt-1993,Williams-etal-2004,Williams-etal-2005} is at the foundation of Einstein's general theory of relativity; therefore, testing the principle is very important. The EP includes three hypotheses: (i) local Lorentz invariance (LLI), (ii) local position invariance (LPI) and (iii) universality of free fall (UFF). Using these three hypotheses Einstein deduced that gravity is a geometric property of spacetime \cite{Will-2006}. One can test both the validity of the EP and of the field equations that determine the geometric structure created by a mass distribution. Below we shall discuss two different ``flavors'' of the Principle,  the weak and the strong forms of the EP that are currently tested in various experiments performed with laboratory test masses and with bodies of astronomical sizes \cite{Williams-etal-2005}.

The \emph{weak form of the EP} (the WEP) states that the gravitational properties of strong and electro-weak interactions obey the EP. In this case the relevant test-body differences are their fractional nuclear-binding differences, their neutron-to-proton ratios, their atomic charges, etc.  Furthermore, the equality of gravitational and inertial masses implies that different neutral massive test bodies will have the same free fall acceleration in an external gravitational field, and therefore in freely falling inertial frames the external gravitational field appears only in the form of a tidal interaction \citep{Singe-1960}. Apart from these tidal corrections, freely falling bodies behave as if external gravity was absent \citep{Anderson-etal-1996}.  

General relativity and other metric theories of gravity assume that the WEP is exact.  However, extensions of the Standard Model of particle physics that contain new macroscopic-range quantum fields predict quantum exchange forces that generically violate the WEP because they couple to generalized ``charges'' rather than to mass/energy as does gravity \cite{Damour-Nordtvedt-1993,Damour-Polyakov-1994,Damour-Piazza-Veneziano-2002}. 
  
Currently, the most accurate results in testing the WEP were reported by ground-based laboratories \cite{Williams-etal-2005,Baessler-etal-1999}. The most recent result \cite{Adelberger_2001,Schlamminger-etal-2007} for the fractional differential acceleration between beryllium and titanium test bodies was given as
%
%
$\Delta a/a=(1.0 \pm 1.4) \times 10^{-13}$. Significant improvements in the tests of the EP are expected from dedicated space-based experiments.

The composition-independence of acceleration rates of various masses 
toward the Earth can be tested to many additional orders of magnitude precision in space-based laboratories, down to levels where some models of the unified theory of quantum  gravity, matter, and energy suggest a possible violation of the EP
\cite{Damour-Nordtvedt-1993,Damour-Polyakov-1994,Damour-Piazza-Veneziano-2002}.  Interestingly, in some scalar-tensor theories, the strength of EP violations and the magnitude of the fifth force mediated by the scalar can be drastically larger in space compared with that on the ground \cite{Mota-Barrow-2004,Khoury-Weltman-2004,Mota-Shaw-2007}, which further justifies a space deployment.  Importantly, many of these theories predict observable violations of the EP at various levels of accuracy ranging from $10^{-13}$ down to $10^{-16}$. Therefore, even a confirmation of no EP-violation will be exceptionally valuable, placing useful constraints on the range of possibilities in the development of a unified physical theory. 

Compared with Earth-based laboratories, experiments in space can benefit from a range of conditions including free-fall and significantly reduced contributions due to seismic, thermal and many other sources of non-gravitational noise (see Appendix~\ref{sec:2}). As a result, there are many experiments proposed to test the EP in space. Below we present only a partial list of these missions.  Furthermore, to illustrate the use of different technologies, we present only the most representative concepts.  

The MicroSCOPE mission\footnote{\label{foot:microscope}Micro-Satellite \`a tra\^in\'ee Compens\'ee pour l'Observation du Principe d'Equivalence (MicroSCOPE), for more details, please see: \tt http://microscope.onera.fr/}  is a room-temperature EP experiment in space relying on electrostatic differential accelerometers \cite{Touboul-Rodrigues-2001}.  The mission is currently under development by CNES\footnote{Centre National d'Etudes Spatiales (CNES) -- the French Space Agency, see website at: {\tt http://www.cnes.fr/}} and ESA, scheduled for launch in 2010.  The design goal is to achieve a differential acceleration accuracy of $10^{-15}$. MicroSCOPE's electrostatic differential accelerometers are based on flight heritage designs from the CHAMP, GRACE and GOCE missions.\footnote{Several gravity missions were recently developed by German National Research Center for Geosciences (GFZ).  Among them are CHAMP (Gravity And Magnetic Field Mission), GRACE (Gravity Recovery And Climate Experiment Mission, together with NASA), and GOCE (Global Ocean Circulation Experiment, together with ESA and other European countries), see {\tt http://www.gfz-potsdam.de/pb1/op/index\_GRAM.html}}

The Principle of Equivalence Measurement (PO\-EM) experiment
\cite{Reasenberg-Phillips-2007} is a ground-based test of the WEP, now under development.  It will be able to detect a violation of the EP with a fractional acceleration accuracy of 5 parts in 10$^{14}$ in a short (few days) experiment and 3 to 10 fold better in a longer experiment.  The experiment makes use of optical distance measurement (by TFG laser gauge \cite{Phillips-Reasenberg-2005}) and  will be  advantageously sensitive to short-range forces with a characteristic length scale of $\lambda < 10$~km. SR-POEM, a POEM-based  proposed room-temperature test of the WEP during a sub-orbital flight on a sounding rocket, was recently also presented \cite{Reasenberg-Phillips-2007-Q2C-II}. It is anticipated to be able to search for a violation of the EP with a single-flight  accuracy of one part in 10$^{16}$. Extension to higher accuracy in an orbital mission is under study. Similarly, the Space Test of Universality of Free Fall (STUFF) \cite{spero-2007} is a recent study of a space-based experiment that relies on optical metrology and proposes to reach an accuracy of one part in 10$^{17}$ in testing the EP in space.

The Quantum Interferometer Test of the Equivalence Principle (QuITE)
\cite{Kasevich-Maleki-2003} is a proposed test of the EP with cold atoms in space. QuITE intends to measure the absolute single axis differential acceleration with accuracy of one part in $10^{16}$, by utilizing two co-located matter wave interferometers with different atomic species.\footnote{Compared to the ground-based conditions, space offers a factor of nearly $10^3$ improvement in the integration times in observation of the free-falling atoms (i.e., progressing from ms to sec).  The longer integration times translate into the accuracy improvements (see discussion in Sec.~\ref{sec:quantum-sensors}).} QuITE will improve the current EP limits set in similar experiments conducted in ground-based laboratory conditions\footnote{Its ground-based analog, called ``Atomic Equivalence Principle Test (AEPT)'', is currently being built at Stanford University. AEPT is designed to reach sensitivity of one part in $10^{15}$.} \cite{Peters-Chung-Chu-2001,Fray-etal-2004} by nearly seven to nine orders of magnitude. 
Similarly, the I.C.E. project\footnote{Interf\'erom\'etrie \`a Source Coh\'erente pour Applications dans l'Espace (I.C.E.), see {\tt http://www.ice-space.fr}} supported by CNES in France aims to develop a high-precision accelerometer based on coherent atomic sources in space \cite{Nyman-etal-2006} with an accurate test of the EP being one of the main objectives.

The Galileo Galilei (GG) mission \cite{Nobili-etal-2007} is an Italian space experiment\footnote{Galileo Galilei (GG) website: {\tt http://eotvos.dm.unipi.it/nobili}} proposed to test the EP at room temperature with accuracy of one part in $10^{17}$.  The key instrument of GG is a differential accelerometer made of weakly-coupled coaxial, concentric test cylinders rapidly spinning around the symmetry axis and sensitive in the plane perpendicular to it. GG  is included in the National Aerospace Plan of the Italian Space Agency (ASI) for implementation in the near future. 

The Satellite Test of Equivalence Principle (STEP) mission \cite{step-2001} is a proposed test of the EP to be conducted from a free-falling platform in space provided by a drag-free spacecraft orbiting the Earth.  STEP will test the composition independence of gravitational acceleration for cryogenically controlled test masses by searching for a violation of the EP with a fractional acceleration accuracy of one part in 10$^{18}$.  As such, this ambitious experiment will be able to test very precisely for the presence of any new non-metric, long range physical interactions.

In its \emph{strong form the EP} (the SEP) is extended to cover the gravitational properties resulting from gravitational energy itself \cite{Williams-etal-2005}.  In other words, it is an assumption about the way that gravity begets gravity, i.e. about the non-linear property of gravitation. Although general relativity assumes that the SEP is exact, alternate metric theories of gravity such as those involving scalar fields, and other extensions of gravity theory, typically violate the SEP. For the SEP case, the relevant test body differences are the fractional contributions to their masses by gravitational self-energy. Because of the extreme weakness of gravity, SEP test bodies must have astronomical sizes. 

Currently, the Earth-Moon-Sun system provides the best solar system arena for testing the SEP. Lunar laser ranging (LLR) experiments involve reflecting laser beams off retroreflector arrays placed on the Moon by the Apollo astronauts and by an unmanned Soviet lander \cite{Williams-etal-2004, Williams-etal-2005}. Recent solutions using LLR data give $(-0.8 \pm 1.3) \times 10^{-13}$ for any possible inequality in the ratios of the gravitational and inertial masses for the Earth and Moon.  This result, in combination with laboratory experiments on the WEP, yields a SEP test of $(-1.8 \pm 1.9) \times 10^{-13}$  that corresponds to the value of the SEP violation parameter of $\eta =(4.0 \pm 4.3) \times 10^{-4}$, where $\eta = 4\beta - \gamma - 3$ and both $\beta$ and $\gamma$ are post-Newtonian parameters \cite{Williams-etal-2005,Williams-2007,PPN}.  

With the new APOLLO\footnote{\label{ref:apollo}The Apache Point Observatory Lunar Laser-ranging Operations (APOLLO) is the new LLR station that was recently  built in New Mexico and successfully initiated operations in 2006.} facility (jointly funded by NASA and NSF, see details in \cite{Murphy-etal-2007,Williams-Turyshev-Murphy-2004}), the LLR science is going through a renaissance.  APOLLO's one-millimeter range precision will translate into order-of-magnitude accuracy improvements in the test of the WEP and SEP (leading to accuracy at the level of ${\Delta a}/{a}\lesssim 1\times10^{-14}$ and $\eta\lesssim 2\times10^{-5}$ correspondingly), in the search for variability of Newton's gravitational constant (see Sec.~\ref{sec:var-fun-const}), and in the test of the gravitational inverse-square law (see Sec.~\ref{sec:inv-sq-law})  
on scales of the Earth-moon distance (anticipated accuracy is $3\times10^{-11}$) \cite{Williams-Turyshev-Murphy-2004}.

The next step in this direction is interplanetary laser ranging \cite{laser-transponders-2006,Chandler-etal-2005,Turyshev-Williams-2007,Merkowitz-etal-2007}, for example, to a lander on Mars. Technology is available to conduct such measurements with a few picoseconds timing precision which could translate into mm-class accuracies achieved in ranging between the Earth and Mars.    The resulting Mars Laser Ranging (MLR) experiment could test the weak and strong forms of the EP with accuracy at the $3\times10^{-15}$ and $2\times10^{-6}$ levels correspondingly, to measure the PPN parameter $\gamma$ (see Sec.~\ref{sec:mod-grav}) with accuracy below the $10^{-6}$ level, and to test gravitational inverse-square law at $\sim2$~AU distances with accuracy of $1\times 10^{-14}$, thereby greatly improving the accuracy of the current tests \cite{Turyshev-Williams-2007}. MLR could also advance research in several areas of science including remote-sensing geodesic and geophysical studies of Mars.

Furthermore, with the recently demonstrated capabilities of reliable laser links over large distances (e.g., tens of millions kilometers) in space \cite{laser-transponders-2006}, there is a strong possibility to improve the accuracy of gravity experiments with precision laser ranging over interplanetary scales \cite{Chandler-etal-2005,Turyshev-Williams-2007,Merkowitz-etal-2007}. Science justification for such an experiment is strong, the required technology is space-qualified and some components have already flown in space. By building MLR, our very best laboratory for gravitational physics will be expanded to interplanetary distances, representing an upgrade in both scale and precision of this promising technique.

The experiments above are examples of the rich opportunities offered by the fundamental physics community to explore the validity of the EP. These experiments could potentially offer up to 5 orders of magnitude improvement over the  accuracy of the current tests of the EP.  Such experiments would dramatically enhance the range of validity for one of the most important physical principles or they could lead to a spectacular discovery.

\subsubsection{\label{sec:var-fun-const}Test of the variation of fundamental constants}

Dirac's 70 year old idea of cosmic variation of physical constants has been revisited with the advent of models unifying the forces of nature based on the symmetry properties of possible extra dimensions, such as the Kaluza-Klein-inspired theories, Brans-Dicke theory, and supersymmetry models.  Alternative theories of gravity \cite{Will-2006} and theories of modified gravity \cite{Bertolami-Paramos-Turyshev-2007} include cosmologically evolving scalar fields that lead to variability of the fundamental constants. Furthermore, it has been  hypothesized that a variation of the cosmological scale factor with epoch could lead to temporal or spatial variation of the physical constants, specifically the gravitational constant, $G$, the fine-structure constant, $\alpha$, and the electron-proton mass ratio ($m_{\rm e}/m_{\rm p}$). 

In general, constraints on the variation of fundamental constants can be derived from a number of gravitational measurements, such as the test of the Universality of Free-Fall, the motion of the planets in the solar system, stellar and galactic evolutions. They are based on the comparison of two time scales, the first (gravitational time) dictated by gravity (ephemeris, stellar ages, etc.), and the second (atomic time) determined by a non-gravitational system (e.g. atomic clocks, etc.) \cite{Canuto-Goldman-1982}. For instance, planetary and spacecraft ranging, neutron star binary observations, paleontological and primordial nucleosynthesis data allow one to constrain the relative variation of $G$ \cite{Uzan-2002}. Many of the corresponding experiments could reach a much higher precision if performed in space. 

A possible variation of Newton's gravitational constant $G$ could be related to the expansion of the universe depending on the cosmological model considered. Variability in $G$ can be tested in space with a much greater precision than on Earth \cite{Williams-etal-2004,Uzan-2002}.  For example, a decreasing gravitational constant, $G$, coupled with angular momentum conservation is expected to increase a planet's semimajor axis, $a$, as $\dot a/a=-\dot G/G$.  The corresponding change in orbital phase grows quadratically with time, providing for strong sensitivity to the effect of $\dot G$. 

Space-based experiments using lunar and planetary ranging measurements currently are the best means to search for very small spatial or temporal gradients in the values of $G$ \cite{Williams-etal-2004,Williams-etal-2005}.  Thus, recent analysis of LLR data strongly limits such variations and constrains a local ($\sim$1~AU) scale expansion of the solar system as $\dot a/a=-\dot G/G =-(5\pm6) \times 10^{-13}$ yr$^{-1}$, including that due to cosmological effects \cite{Williams-etal-2007,Williams-2007}. Interestingly, the achieved accuracy in $\dot G/G$ implies that, if this rate is representative of our cosmic history, then $G$ has changed by less than 1\% over the 13.4 Gyr age of the universe.  

The ever-extending LLR data set and increase in the accuracy of lunar ranging (i.e., APOLLO) could lead to significant improvements in the search for variability of Newton's gravitational constant; an accuracy at the level of $\dot G/G\sim 1\times10^{-14}~{\rm yr}^{-1}$ is feasible with LLR \cite{Turyshev-Williams-2007}. High-accuracy timing measurements of binary and double pulsars could also provide a good test of the variability of the gravitational constant \cite{Nordtvedt-2002,Kramer-etal-2006}.  

The current limits on the evolution of $\alpha$ are established by
laboratory measurements, studies of the abundances of radioactive isotopes and
those of fluctuations in the cosmic microwave background, as well as other cosmological constraints (for review see \cite{Uzan-2002}).  Laboratory experiments are based on the comparison either of different atomic clocks or of atomic clocks with ultra-stable oscillators. They also have the advantage of being more reliable and reproducible, thus allowing better control of the systematics and better statistics compared with other methods. Their evident drawback is their short time scales, fixed by the fractional stability of the least precise standards. These time scales usually are of order of a month to a year so that the obtained constraints are restricted to the instantaneous variation today.  However, the shortness of the time scales is compensated by a much higher experimental sensitivity. 

There is a connection between the variation of the fundamental constants and a violation of the EP; in fact, the former almost always implies the latter.\footnote{Note, the converse is not necessarily true:  the EP may be violated without any observable variation of fundamental constants.} For example, should there be an ultra-light scalar particle, its existence would lead to variability of the fundamental constants, such as $\alpha$ and $m_{\rm e}/m_{\rm p}$. Because  masses of nucleons are $\alpha$-dependent, by coupling to nucleons this particle would mediate an isotope-dependent long-range force \cite{Damour-Polyakov-1994,Damour-2002,Dvali-Zaldarriaga-2002,Uzan-2002,Dent-2007}. The strength of the coupling is within a few of orders of magnitude from the existing experimental bounds for such forces; thus, the new force can be potentially measured in precision tests of the EP. Therefore, the existence of a new interaction mediated by a massless (or very low-mass) time-varying scalar field would lead to both the variation of the fundamental constants and violation of the WEP, ultimately resulting in observable deviations from general relativity.

Following the arguments above, for macroscopic bodies, one expects that their masses depend on all the coupling constants of the four known fundamental interactions, which has profound consequences concerning the motion of a body. In particular, because the $\alpha$-dependence is {\it a priori} composition-dependent, any variation of the fundamental constants will entail a violation of the universality of free fall \cite{Uzan-2002}. This allows one to compare the ability of two classes of experiments -- clock-based and EP-testing ones -- to search for variation of the parameter $\alpha$ in a model-independent way \cite{Nordtvedt-2002}. EP experiments have been superior performers. Thus, analysis of the frequency ratio of the 282-nm $^{199}$Hg$^+$ optical clock transition to the ground state hyperfine splitting in $^{133}$Cs was recently used to place a limit on its fractional variation of $\dot\alpha/\alpha \leq 1.3\times 10^{-16}~{\rm yr}^{-1}$ \cite{Fortier-etal-2007}. At the same time, the current accuracy of the EP tests \cite{Williams-etal-2005} already constrains the variation as $\Delta \alpha/\alpha \leq 10^{-10}\Delta U/c^2$, where $\Delta U$ is the change in the gravity potential. Therefore, for ground-based experiments (for which the variability in the gravitational potential is due to the orbital motion of the Earth) in one year the quantity $U_{\tt sun}/c^2$ varies by $1.66\times10^{-10}$, so a ground-based clock experiment must therefore be able to measure fractional frequency shifts between clocks to a precision of a part in $10^{20}$ in order to compete with EP experiments on the ground \cite{Nordtvedt-2002}. 


On the other hand, sending atomic clocks on a spacecraft to within a few solar radii of the Sun where the gravitational potential grows to $10^{-6} c^2$ could, however, be a competitive experiment if the relative frequencies of different on-board clocks could be measured to a precision better than a part in $10^{16}$. Such an experiment would allow for a direct measurement of any $\alpha$-variation, thus further motivating the development of  space-qualified clocks. With their accuracy surpassing the $10^{-17}$ level in the near future, optical clocks may be able to provide the needed capabilities  to directly test the variability of the fine-structure constant (see Sec.~\ref{sec:clocks} for details). 

{SpaceTime} is a proposed atomic-clock experiment designed to search
for a variation of the fine-structure constant with a detection sensitivity of $\dot\alpha/\alpha\sim 10^{-20}~{\rm yr}^{-1}$ and will be carried out on a spacecraft that flies to within six solar radii of the sun \cite{Maleki-Prestage-2007}. The test relies on an instrument utilizing a tri-clock assembly that consists of three trapped-ion clocks based on mercury, cadmium, and ytterbium ions that are placed in the same vacuum, thermal and magnetic field environment.  Such a configuration allows for a differential measurement of the frequency of the clocks, and the cancellation of perturbations common to the three. For alkali atoms, the sensitivity of different clocks, based on atoms of different $Z$, to a change in the fine structure constant display specific signatures.  In particular, the Casimir correction factor, $F(\alpha Z)$, leads to the differential sensitivity in the alkali microwave hyperfine clock transition frequencies. As a result, different atomic systems with different $Z$ display different frequency dependencies on a variation of $\alpha$ through the $\alpha Z$ dependent terms.  A direct test for a time variation of $\alpha$ can then be devised through a comparison of two clocks, based on two atomic species with different atomic number, $Z$.  This is a key feature of the SpaceTime instrument that in conjunction with the individual sensitivity of each atomic species to an $\alpha$-variation, can produce clear and unambiguous results. Observation of any frequency drift between the three pairs of the clocks in response to the change in gravitational potential, as the tri-clock instrument approaches the sun, would signal a variation in $\alpha$.

Clearly a solar fly-by on a highly-eccentric trajectory with very accurate clocks and inertial sensors makes for a compelling relativity test.  A potential use of highly-accurate optical clocks (see Sec.~\ref{sec:clocks}) in such an experiment would likely lead to additional accuracy improvement in the tests of $\alpha$ and $m_{\rm e}/m_{\rm p}$, thereby providing a good justification for space deployment \cite{Schiller-etal-2006}. The resulting space-based laboratory experiment could lead to an important discovery. 

\subsubsection{\label{sec:inv-sq-law}Search for new physics via tests of the gravitational inverse square law}  

Many modern theories of gravity, including string, supersymmetry, and brane-world theories, have suggested that new physical interactions will appear at short ranges.  This may happen, in particular, because at sub-millimeter distances new dimensions can exist, thereby changing the gravitational inverse-square law \cite{ADD-1998,ADD-1999} (for review of experiments, see \cite{Adelberger-Heckel-Nelson-2003}). Similar forces that act at short distances are predicted in supersymmetric theories with weak scale compactifications \cite{AnDD-1998}, in some theories with very low energy
supersymmetry breaking \cite{Dimopoulos-Giudice-1998}, and also in theories of
very low quantum gravity scale \cite{Sundrum-1999,Dvali-etal-1998}. These multiple predictions provide strong motivation for experiments that would test for possible deviations from Newton's gravitational inverse-square law at very short distances, notably on ranges from 1 mm to 1 $\mu$m.

Recent ground-based torsion-balance experiments \cite{Kapner-etal-2007} tested the gravitational inverse-square law at separations between 9.53 mm and $55~\mu$m, probing distances less than the dark-energy length scale $\lambda_d =\sqrt[4]{\hbar c/ u_d}\approx 85~\mu$m, with energy density $u_d\approx3.8~{\rm keV/cm}^3$.  It was found that the inverse-square law holds down to a length scale of $56~\mu$m and that an extra dimension must have a size less than $44~\mu$m (similar results were obtained by \cite{Tu-etal-2007}).  These results are important, as they signify the fact that modern experiments reached the level at which dark-energy physics can be tested in a laboratory setting; they also provided a new set of constraints on new forces \cite{Adelberger-etal-2007}, making such experiments very relevant and competitive with particle physics research.  In addition, recent laboratory experiments testing the Newton's second law for small accelerations \cite{Schlamminger-etal-2007,Gundlach-etal-2007} also provided useful constraints relevant to understanding several current astrophysical puzzles.  

Sensitive experiments searching for weak forces invariably require soft suspension for the measurement degree of freedom.  A promising soft suspension with low dissipation is superconducting magnetic levitation.  Levitation in 1-{\sl g}, however, requires a large magnetic field, which tends to couple to the measurement degree of freedom through metrology errors and coil non-linearity, and stiffen the mode.  The high magnetic field will also make suspension more dissipative.  The situation improves dramatically in space.  The {\sl g}-level is reduced by five to six orders of magnitude, so the test masses can be supported with weaker magnetic springs, permitting the realization of both the lowest resonance frequency and lowest dissipation.  The microgravity conditions also allow for an improved design of the null experiment, free from the geometric constraints of the torsion balance.

The Inverse-Square Law Experiment in Space (ISLES) is a proposed experiment whose objective is to perform a highly accurate test of Newton's gravitational law in space \cite{Paik-etal-2007}. ISLES combines the advantages of the microgravity environment  with superconducting accelerometer technology to improve the current ground-based limits in the strength of violation \cite{Chiaverini-etal-2003} by four to six orders of magnitude in the range below $100~\mu$m.  The experiment will be sensitive enough to probe large extra dimensions down to $5~\mu$m and also to probe the existence of the axion\footnote{The axion is a hypothetical elementary particle postulated by Peccei-Quinn theory in 1977 to resolve the strong-CP problem in quantum chromodynamics (QCD); see details in Ref.~\cite{Peccei-Quinn-1977}.} which, if it exists, is expected to violate the inverse-square law in the range accessible by ISLES.

The recent theoretical ideas concerning new particles and new dimensions have reshaped the way we think about the universe. Thus, should the next generation of experiments detects a force violating the inverse-square law, such a discovery would imply the existence of either an extra spatial dimension, or a massive graviton, or the presence of a new fundamental interaction \cite{Adelberger-Heckel-Hoyle-2005}.   

While most attention has focused on the behavior of gravity at short distances, it is possible that tiny deviations from the inverse-square law occur at much larger distances. In fact, there is a possibility that non-compact extra dimensions could produce such deviations at astronomical distances \cite{Dvali-Gruzinov-Zaldarriaga-2003} (for discussion see Sec.~\ref{sec:mod-grav}).

By far the most stringent constraints on a test of the inverse-square law to date come from very precise measurements of the Moon's orbit about the Earth. Even though the Moon's orbit has a mean radius of 384,000 km, the models agree with the data at the level of 4 mm! As a result, analysis of the LLR data tests the gravitational inverse-square law to $3\times10^{-11}$ of the gravitational field strength on scales of the Earth-moon distance \cite{Williams-Turyshev-Murphy-2004}.

Interplanetary laser ranging could provide conditions that are needed to improve the tests of the inverse-square law on the interplanetary scales \cite{Turyshev-Williams-2007}. MLR could be used to perform such an experiment that could reach the accuracy of $1\times 10^{-14}$ at 2~AU distances, thereby improving the current tests  by several orders of magnitude. 

Although most of the modern experiments do not show disagreements with Newton's law, there are puzzles that require further investigation. The radiometric tracking data received from the Pioneer 10 and 11 spacecraft at heliocentric distances between 20 and 70 AU has consistently indicated the presence of a small, anomalous, Doppler drift in the spacecraft carrier frequency. The drift can be interpreted as due to a constant sunward acceleration of $a_{\rm P}  = (8.74 \pm 1.33)\times 10^{-10}~{\rm m/s}^2$ for each particular craft \cite{PA-papers}. This apparent violation of the inverse-square law has become known as the Pioneer anomaly.

The possibility that the anomalous behavior will continue to defy attempts at a conventional explanation has resulted in a growing discussion about the origin of the discovered effect, including suggestions for new physics mechanisms \cite{PA-theory} and proposals for a dedicated deep space experiment \cite{PA-work}.\footnote{For details, please consult the web-page of the Pioneer Explorer Collaboration at the International Space Science Institute (ISSI), Bern, Switzerland, {\tt http://www.issi.unibe.ch/teams/Pioneer/}} A recently initiated investigation of the anomalous signal using the entire record of the Pioneer spacecraft telemetry files in conjunction with the analysis of a much extended Pioneer Doppler data may soon reveal 
the origin of the  anomaly \cite{pio-data}. 

Besides the Pioneer anomaly, there are other intriguing puzzles in the solar system dynamics still awaiting a proper explanation, notably the so-called `fly-by anomaly' \cite{fly-by}, that occurred during Earth gravity assists performed by several interplanetary spacecraft.

\subsubsection{\label{sec:mod-grav}Tests of alternative and modified gravity theories
with gravitational experiments in the solar system}

Given the immense challenge posed by the unexpected discovery of the accelerated expansion of the universe, it is important to explore every option to explain and probe the underlying physics.  Theoretical efforts in this area offer a rich spectrum of new ideas, some of them are discussed below, that can be tested by experiment.

Motivated by the dark energy and dark matter problems, long-distance gravity modification is one of the radical proposals that has recently gained attention \cite{Deffayet-Dvali-Gabadadze-2002}.  Theories that modify gravity at cosmological distances exhibit a strong coupling phenomenon of extra graviton polarizations \cite{Deffayet-etal-2002,Dvali-2006}. This strong coupling phenomenon plays an important role for this class of theories in allowing them to agree with solar system constraints.  In particular, the ``brane-induced gravity'' model \cite{Dvali-Gabadadze-Porrati-2003} provides a new and interesting way of modifying gravity at large distances to produce an accelerated expansion of the universe, without the need for a non-vanishing cosmological constant \cite{Deffayet-2001}. One of the peculiarities of this model is the way one recovers the usual gravitational interaction at small (i.e. non-cosmological) distances, motivating precision tests of gravity on solar system scales \cite{Sahni-Shtanov-2006,Bekenstein-Magueijo-2006}.   

The Eddington parameter $\gamma$, whose value in general relativity is unity, is perhaps the most fundamental parameterized-post-Newtonian (PPN) parameter \cite{PPN,Will-2006}, in that $\frac{1}{2}(1-\gamma)$ is a measure, for example, of the fractional strength of the scalar gravity interaction in scalar-tensor theories of gravity \cite{Damour-Nordtvedt-1993}.  Currently, the most precise value for this parameter, $\gamma -1 = (2.1\pm2.3)\times10^{-5}$, was obtained using radio-metric tracking data received from the Cassini spacecraft \cite{Bertotti-Iess-Tortora-2003} during a solar conjunction experiment. This accuracy approaches the region where multiple tensor-scalar gravity models, consistent with the recent cosmological observations \cite{Spergel-etal-2006}, predict a lower bound for the present value of this parameter at the level of $(1-\gamma) \sim 10^{-6}-10^{-7}$ \cite{Damour-Nordtvedt-1993,Damour-Polyakov-1994,Damour-Piazza-Veneziano-2002,Damour-Esposito-Farese-1996}.  Therefore, improving the measurement of this parameter\footnote{In addition, any experiment pushing the present upper bounds on another Eddington  parameter $\beta$, i.e. $\beta - 1 = (0.9 \pm 1.1) \times 10^{-4}$ from \citep{Williams-etal-2004,Williams-etal-2005}, will also be of interest.} would provide crucial information to separate modern scalar-tensor theories of gravity from general relativity, probe possible ways for gravity quantization, and test modern theories of cosmological evolution.

Interplanetary laser ranging could lead to a significant improvement in the accuracy of the parameter $\gamma$. Thus, precision ranging between the Earth and a lander on Mars during solar conjunctions may offer a suitable opportunity (i.e., MLR).\footnote{In addition to Mars, a Mercury lander \cite{Anderson-etal-1997} equipped with a laser ranging transponder would be very interesting as it would probe a stronger gravity regime while providing measurements that will not be affected by the dynamical noise from the asteroids \cite{Ken_asteroids97,Konopliv_etal_2005}.} If the lander were to be equipped with a laser transponder capable of reaching a precision of 1~cm, a measurement of $\gamma$ with accuracy of 1 part in 10$^6$ is possible. To reach accuracies beyond this level one must rely on a dedicated space experiment \cite{Turyshev-Williams-2007}.

The Gravitational Time Delay Mission (GTDM) \cite{Ashby-Bender-2006,Bender-etal-2006} proposes to use laser ranging between two drag-free spacecraft (with spurious acceleration levels below $1.3 \times  10^{-13}~{\rm m/s}^2/\sqrt{\rm Hz}$ at  $0.4~\mu$Hz) to accurately measure the Shapiro time delay for laser beams passing near the Sun. One spacecraft would be kept at the L1 Lagrange point of the Earth-Sun system with the other one being placed on a 3:2 Earth-resonant, LATOR-type, orbit (see \cite{Turyshev-etal-2006} for details). A high-stability frequency standard ($\delta f/f\lesssim 1 \times  10^{-13}~1/\sqrt{\rm Hz}$ at $0.4~\mu$Hz) located on the L1 spacecraft permits accurate measurement of the time delay. If requirements on the performance of the disturbance compensation system, the timing transfer process, and high-accuracy orbit determination are successfully addressed \cite{Bender-etal-2006}, then determination of the time delay of interplanetary signals to 0.5 ps precision in terms of the instantaneous clock frequency could lead to an accuracy of 2 parts in $10^{8}$ in measuring the parameter $\gamma$.

The Laser Astrometric Test of Relativity (LATOR) \cite{Turyshev-etal-2006,hellings_2005} proposes to measure the parameter $\gamma$ with accuracy of a part in 10$^9$, which is a factor of 30,000 beyond the currently best Cassini's 2003 result \cite{Bertotti-Iess-Tortora-2003}.  The key element of LATOR is a geometric redundancy provided by the long-baseline optical interferometry and interplanetary laser ranging. 
By using a combination of independent time-series of gravitational deflection of light in the immediate proximity to the Sun, along with measurements of the Shapiro time delay on interplanetary scales (to a precision better than 0.01 picoradians and 3 mm, respectively), LATOR will significantly improve our knowledge of relativistic gravity and cosmology. LATOR's primary measurement, precise observation of the non-Euclidean geometry of a light triangle that surrounds the Sun, pushes to unprecedented accuracy the search for cosmologically relevant scalar-tensor theories of gravity by looking for a remnant scalar field in today's solar system.  LATOR could lead to very robust advances in the tests of fundamental physics -- it could discover a violation or extension of general relativity or reveal the presence of an additional long range interaction.

If implemented, the missions discussed above could significantly advance research in fundamental physics; however, due to lack of dedicated support none of these experiments could currently be performed in the U.S. The recent trend in allowing ``laboratory'' fundamental physics to compete for funding together with ``observational'' disciplines of space sciences may change the situation and could help NASA to enhance its ongoing space research efforts, so that the space agency will be doing the best possible science in space -- a truly noble objective. 

\subsection{\label{sec:2.3} Detection and study of gravitational waves}

Gravitational waves, a key prediction of Einstein's general theory of relativity have not yet been directly detected.  The only indirect evidence for existence of gravitational waves came from binary pulsar investigation, the discovery that led to the 1993 Nobel Prize in physics.

Gravitational wave observatories in space will provide insight into the structure and dynamics of space and time.  They will also be able to detect the signatures of cosmic superstrings or phase transitions in the early universe and contribute to the study of the dark energy that dominates the evolution of the universe. Space-based gravitational wave observatories, such as the planned Laser Interferometer Space Antenna (LISA)\footnote{\label{lisa}Laser Interferometric Space Antenna (LISA), an international space mission whose development is currently funded through a collaborative agreement between NASA and ESA, see websites: {\tt http://lisa.nasa.gov/},  {\tt http://www.lisa-science.org/} and resources therein.},  will offer access to a range of the gravitational wave frequency spectrum that is not accessible on the Earth.  
%
%
LISA promises to open a completely new window into the heart of the most energetic processes in the universe, with consequences fundamental to both physics and astronomy. LISA is a joint NASA-ESA mission that expects to detect gravitational waves from the merger of massive black holes in the centers of galaxies or stellar clusters at cosmological distances, and from stellar mass compact objects as they orbit and fall into massive black holes. LISA will measure the signals from close binaries of white dwarfs, neutron stars, or stellar mass black holes in the Milky Way and nearby galaxies. 

LISA will consist of an array of three spacecraft orbiting the sun, each separated
from its neighbor by about 5 million kilometers. Laser beams will be used to measure the minute changes in distance between the spacecraft induced by passing gravitational waves. 
For this purpose, the spacecraft have to be drag-free, a requirement common for many fundamental physics missions.  The preparatory LISA Pathfinder\footnote{\label{lpf}LISA Pathfinder is a technology demonstration for the future LISA mission, see {\tt http://sci.esa.int/science-e/www/area/ index.cfm?fareaid=40/}} mission is aimed at demonstrating the ability to achieve free-fall conditions at the required levels of accuracy.

Studies of gravitational waves can provide potentially powerful insight for large-distance modified gravity theories.  The main reason is that in all such theories the graviton carries 3 extra polarizations (for details, see \cite{Dvali-1997}), which follows from similar properties \cite{vanDam-Veltman-1970} of the massive graviton \cite{Fierz-Pauli-1939}.  This is in contrast to the normal massless graviton in general relativity that carries only two helicities.  In addition, in such theories the dispersion relation of the graviton is no longer that of a massless spin-2 particle, but rather acquires non-trivial frequency dependence \cite{Dvali-1997}.  As a result, both emission and propagation properties of the gravitational waves are altered in modified-gravity theories. LISA and future gravitational wave missions will be able to address these important questions and provide insight needed to explore these possibilities.

Although of extreme importance, research in gravitational waves does not enjoy stable funding. NASA support to research in this area is conducted via the {\it Beyond Einstein} Program\footnote{\label{be-program}NASA Beyond Einstein Program: {\tt http://universe.nasa.gov/}} that was approved by the U.S. Congress in 2004 and is managed by the Astrophysics Division of NASA's Science Mission Directorate. The program was recently assessed by the Beyond Einstein Program Assessment Committee (BEPAC) formed by the Space Studies Board and the Board on Physics and Astronomy of the National Academy of Sciences (NAS) with the purpose to assess the five proposed Beyond Einstein missions (Constellation-X, LISA, Joint Dark Energy Mission (JDEM), Inflation Probe, and Black Hole Finder probe) and to recommend which of these five missions should be developed and launched first.\footnote{For details, please see: {\tt http://www7.nationalacademies.org/ ssb/BeyondEinsteinPublic.html }} 
In it's recently released Report\footnote{\label{bepac-report}``NASA's Beyond Einstein Program: An Architecture for Implementation'', Committee on NASA's Einstein Program: An
Architecture for Implementation, National Research Council, 
for details, please see: {\tt http://www.nap.edu/catalog/12006.html }}, BEPAC recommended NASA to immediately proceed with the development of the JDEM mission, while also investing additional funds in LISA technology development and risk reduction. 

Although, the NAC's assessment effectively postpones LISA's launch towards the end of the next decade, it urges NASA to develop LISA-enabling technologies, many of which are common to other fundamental physics missions.  As such, BEPAC recommendations will have major impact on the entire field of space-based research in fundamental physics in the next decade and beyond. 

\subsection{\label{sec:2.4} Precision research in cosmology}

The current model of the universe includes critical assumptions, such as an inflationary epoch in primordial times, and peculiar settings, such as the fine tuning in the hierarchy problem, that call for a deeper theoretical framework.  In addition, the very serious vacuum and/or dark energy problems and the related cosmological phase transitions lead researchers to areas beyond general relativity and standard quantum field theory.  Observations of the early universe are an important tool in constraining physics beyond the standard model and quantum gravity.  This work led to the discovery of the fluctuations in the Cosmic Microwave Background (CMB) made by the NASA COBE mission -- an experiment that revolutionized the entire field of cosmology and led to the 2006 Nobel Prize in physics.

The 2005 report by the CMB Task Force\footnote{See the webpage of the Task Force of CMB at the NSF: {\tt http://www.nsf.gov/mps/ast/tfcr.jsp}  and also ``Report from the Task Force on CMB Research (TFCR),'' July 11, 2005 {\tt http://www.nsf.gov/mps/ast/tfcr\_final\_report.pdf}}  identified two types of observations that are critical for cosmological research: (1) study of the polarization of the CMB anisotropies and (2) direct detection of primordial gravitational waves with second- or third-generation missions.  Some of the relevant observations can be made by space missions inspired by the astronomy community.  Other observations will be made by space missions inspired by the fundamental physics community.  It seems to be quite natural that the former missions should be under the purview of astronomy, while the latter missions should fall under the purview of fundamental physics.  

The recently discovered baryon acoustic oscillations \cite{Eisenstein-etal-2005} together with the CMB and supernovae data provide additional very important constraints for possible models and scenarios. The 2006 Report from the Dark Energy Task Force (DETF)\footnote{Dark Energy Task Force (DETF) website at the NSF is at {\tt http://www.nsf.gov/mps/ast/detf.jsp}} explicitly mentions the significance of various tests of general relativity, especially as they relate to dark energy \cite{DETF-report-2006}.  The report also highlights the synergy between observational and experimental methods to benefit modern research in cosmology. Just as dark-energy science has far-reaching implications for other fields of physics, advances and discoveries made in laboratory fundamental physics may point the way toward understanding the nature of dark energy.  For instance, such a pointer could come from observing any evidence of a failure of general relativity.
 
The strong coupling phenomenon (discussed in Section~\ref{sec:2.1}) makes modified gravity theories predictive and potentially testable at scales that are much shorter than the current cosmological horizon.  Because of the key role that non-linearities play in relativistic cosmology, namely those of its scalar sector, their presence leads to potentially observable effects in gravitational studies within our solar system.  Thus, it is possible to test some features of cosmological theories in space-based experiments performed on spacecraft-accessible distances \cite{Dvali-Gruzinov-Zaldarriaga-2003}.

There is a profound connection between cosmology and possible Lorentz symmetry violation \cite{Kostelecky-2004,Dvali-Pujolas-Redi-2007} (see also Section~\ref{sec:2.2}). Spontaneous breaking of the Lorentz symmetry implies that there exists an order parameter with a non-zero expectation value that is responsible for the effect. For spontaneous Lorentz symmetry breaking one usually assumes that sources other than the familiar matter density are responsible for such a violation.  However, if Lorentz symmetry is broken by an extra source, the latter must also affect the cosmological background.  Therefore, in order to identify the mechanism of such a violation, one has to look for traces of similar symmetry breaking in cosmology, for instance, in the CMB data.\footnote{Analyses of the CMB for Lorentz violation have already begun \cite{Feng-etal-2006,Kostelecky-Mewes-2007}. This provides a systematic classification of all operators for Lorentz violation and uses polarimetric observations of the cosmic microwave background to search for associated effects. Lorentz symmetry violation can also have important implication for cosmology via CPT violation and baryogenesis \cite{Bertolami-etal-1997,Li-etal-2007}.} In other words, should a violation of the Lorentz symmetry be discovered in experiments but not supported by observational cosmology data, such a discrepancy would indicate the existence of a novel source of symmetry breaking.  This source would affect the dispersion relation of particles and the performance of the local clocks, but leave no imprint on the cosmological metric.  Such a possibility emphasizes the importance of a comprehensive program to investigate all possible mechanisms of breaking of the Lorentz symmetry, including those accessible by experiments conducted in space-based laboratories.

Because of the recent important discoveries, the area of observational cosmology is receiving some limited multi-agency support from NASA, DOE, and NSF.  However, no support is available for laboratory experiments in this discipline. NASA's support for research in cosmology comes through the {Beyond Einstein Program}, but it is limited to observational aspects, providing essentially no support to relevant solar-system laboratory experiments.

\subsection{\label{sec:2.5} Space-based efforts in astroparticle physics}

Astroparticle physics touches the foundation of our understanding of the matter content of the universe.  One can use cosmic rays, high-energy photons, and neutrinos to test the fundamental laws of nature at energies well beyond the reach of terrestrial experiments.  The results can play an important role in the future development of the fundamental theory of elementary particles.  Many observations require going to space because the atmosphere stops the cosmic messengers, such as X-rays.   In other cases, for example, in the case of ultrahigh-energy cosmic rays, the advantage the space missions offer is in observing a large segment of the Earth's atmosphere, which can be used as part of the detector.  Below we will discuss some examples of how the astroparticle physics can benefit from going to space. 

\subsubsection{Detection of ultrahigh-energy cosmic rays and neutrinos from space}

Ultrahigh-energy cosmic rays (UHECR), with energies in excess of 10$^{20}$ eV and beyond, have been observed by the AGASA\footnote{The Akeno Giant Air Shower Array (AGASA) experiment, please see website: {\tt
http://www-akeno.icrr.u-tokyo.ac.jp/AGASA/}}, HiRes\footnote{The High
Resolution Fly's Eye (HiRes) experiment, please see website: {\tt
http://hires.phys.columbia.edu/}}, and Pierre Auger\footnote{The Pierre Auger Cosmic Ray Observatory, please see  website: {\tt http://www.auger.org/}} experiments.  Understanding the origin and propagation of these cosmic rays may provide a key to fundamental laws of physics at the highest energy scales, several orders of magnitude beyond the reach of particle accelerators.  The propagation of cosmic rays through space will test both the fundamental symmetry of
space-time and the Lorentz invariance.   

Detection of UHE neutrinos will mark the beginning of a new era in astronomy and will allow mapping of the most extreme objects in the universe, such as super-massive black holes, active galactic nuclei, and, possibly, cosmic strings and other topological defects.   

By comparing the rates of upgoing and downgoing neutrino-initiated air
showers, one can measure the neutrino-nucleon cross section at the
center-of-mass energy as high as $10^6$~GeV, orders of magnitude beyond  the reach of collider experiments~\cite{Kusenko:2001gj}.  There are differing theoretical predictions for this cross section; its measurement can probe fundamental physics at the highest scales~\cite{gzk,Kusenko:2001gj}. 
 
Space-based instruments, such as the proposed EUSO\footnote{The Extreme
Universe Space Observatory (EUSO), please see website:  {\tt
http://www.euso-mission.org/}}  and OWL\footnote{The Orbiting Wide-angle Light collectors (OWL), please see website: {\tt http://owl.gsfc.nasa.gov/}}, can use a large segment of the Earth's atmosphere as a medium for detecting cosmic rays
and neutrinos.  These observations can test the fundamental laws of physics at the highest energy frontier.  They also provide information about the most extreme objects in the universe, such as supermassive black holes.

\subsubsection{Identifying the dark-matter particles by their properties}

There is now overwhelming evidence that most of the matter in the universe is not made of ordinary atoms, but, rather, of new, yet undiscovered, particles (see \cite{Bertone-Hooper-Silk-2005} for review).  The evidence for dark matter is based on several independent observations, including the anisotropies of the cosmic microwave background radiation, gravitational lensing, optical observations of the galactic rotation curves, and x-ray observations of clusters.  None of the Standard Model particles can be dark matter.  Hence, the identification of dark matter will be a discovery of new physics beyond the Standard Model.

One of the most popular theories for physics beyond the Standard Model is supersymmetry (SUSY).  A class of supersymmetric extensions of the Standard Model predicts dark matter in the form of either the lightest supersymmetric particles (LSP) (see \cite{Jungman-Kamionkowski-Griest-2005} for review),  or SUSY Q-balls \cite{Kusenko-1997}. Another theoretically appealing possibility is dark matter in the form of axions \cite{Peccei-Quinn-1977}.   An axion is a very weakly interacting field that accompanies the Peccei--Quinn solution of the strong CP problem.\footnote{The existence of the axion is suggested by models attempting to solve symmetry problems of the Standard Model \cite{Peccei-Quinn-1977}.  The axion would violate the $1/r^2$ law of gravity at short distances and would thus be detectable experimentally.}   There are several other dark-matter candidates that are well motivated by theoretical reasoning.

The right-handed or sterile neutrinos can be the cosmological dark matter
\cite{Dodelson-Widrow-1994}.   The existence of such right-handed neutrino
states is implied by the discovery of the active neutrino masses.  Although it
is not impossible to explain the neutrino masses otherwise, most models
introduce gauge singlet fermions that give the neutrinos their masses via
mixing.  If one of these right-handed states has a mass in the range of $\sim 1-50$~keV, it can be the dark matter.\footnote{LSND experiment claimed \cite{lsnd} to observe a sterile neutrino with a much smaller mass ($m\sim{\rm eV}$)  and a much larger mixing angle ($\sin^2\theta \sim 10^{-1}$) than those needed for dark matter
($m\sim{\rm keV}$, $\sin^2\theta \sim 10^{-9}$). The LSND neutrino would pose some
serious problems for cosmology. Recent results from the MiniBooNE experiment \cite{miniboone} have refuted the LSND claim.}  Several indirect astrophysical clues support this hypothesis.  Indeed, if sterile neutrinos exist, they can explain the
long-standing puzzle of pulsar velocities \cite{Kusenko-Segre-1997}.   In
addition, the x-rays produced in decays of the relic neutrinos could increase
the ionization of the primordial gas and can catalyze the formation of molecular
hydrogen at redshifts as high as 100.  Since molecular hydrogen is an important
cooling agent, its increased abundance could play an important role in the
foremation of the first stars \cite{Biermann-Kusenko-2006}.  Sterile neutrinos
can also help the formation of supermassive black holes in the early universe, as well as
explain the matter-antimatter asymmetry \cite{Akhmedov-Rubakov-Smirnov-1998}.  
The consensus of these indirect observational hints makes a stronger case for
the sterile dark matter \cite{Kusenko-2007}.

Depending on the properties of dark-matter particles, they can be identified by
one of several techniques.  In the case of LSP, their annihilations in the
center of our galaxy can produce gamma rays.  Thus, the search for gamma rays
from the Galactic Center with instruments such as GLAST\footnote{The Gamma-ray
Large Area Space Telescope (GLAST), please see website: {\tt
http://glast.gsfc.nasa.gov/}}  may lead to the discovery of this form of dark
matter \cite{Moriselli-etal-2002}.  The spectrum, the flux, and the
distribution of these gamma-rays can be used to distinguish the SUSY signal
from the alternatives.  The LSP annihilations are also expected to produce an
identifiable flux of antiprotons and antideuterons \cite{Baer-Profumo-2005}.
Superheavy dark matter can be discovered by the proposed EUSO or OWL if the
heavy particles decay producing the UHECRs, as suggested by some theories.

If dark matter is made up of sterile neutrinos with mass in the keV
range~\cite{Dodelson-Widrow-1994}, their decays into the lighter left-handed
neutrinos and x-rays offer an opportunity to discover dark matter using a
high-resolution x-ray spectrometer. Since this decay is a two-body process, the
decay photons produce a narrow spectral line, Doppler-broadened due to the
motion of dark-matter particles. For the most plausible masses, one expects a
line between 1 and 50~keV with width of about 1 eV.  To detect this line and
distinguish it from the gaseous lines, one needs an instrument with a good
energy resolution. Current limits are based on the observations of Chandra and
XMM-Newton~\cite{x-rays}. A dedicated search by the Suzaku telescope is under
way. Further ideas for space-based experiments searching for sterile neutrinos
are being investigated.  In particular, the recently proposed x-ray telescope
in space, EDGE \footnote{The Explorer of Diffuse emission and Gamma-ray burst
Explosions (EDGE), see mission's website at:
{\tt http://projects.iasf-roma.inaf.it/edge/EdgeOverview.htm}}, would be able to
search for dark matter in the form of sterile neutrinos.  

Ongoing efforts in the search for dark matter particles include a number of underground detectors, as well as GLAST.   The search for dark matter must cover a number of avenues because different candidate particles have different interactions. NASA has a unique opportunity to initiate and lead the work in this very important area of space-based research, by funding missions such as EDGE and others, which could potentially result in a major discovery. Similar opportunities for a significant discovery exist in other areas of astroparticle physics.

\subsection{\label{sec:2.2} Search for physics beyond the Standard Model with
space-based experiments}

The Standard Model coupled to general relativity is thought to be the
effective low-energy limit of an underlying fundamental theory that unifies gravity and gravity and particle physics at the Planck scale. This underlying theory may well include Lorentz violation \cite{Colladay-Kostelecky-1997,Coleman-Glashow-1997} which could be detectable in space-based experiments \cite{Kostelecky-Potting-1995}.   If one takes the Standard Model and adds appropriate terms that involve operators for Lorentz invariance violation \cite{Stecker-Glashow-2001}, the result is the Standard-Model Extension (SME), which has provided a phenomenological framework for testing Lorentz-invariance
\cite{Kostelecky-Samuel-1991,Kostelecky-Samuel-1989}, and also suggested a number of new tests of relativistic gravity in the solar system \cite{Bailey-Kostelecky-1989}. Compared with their ground-based analogs, space-based experiments in this area can provide improvements by as much as six orders of magnitude.  Several general reviews of the SME and corresponding efforts are available (for review, see \cite{review-Kostelecky-1989}).   Recent studies of the ``aether theories'' \cite{Foster-Jacobson-2005} have shown that these models are naturally compatible with general relativity \cite{Will-2006}, but predict several non-vanishing Lorentz-violation parameters that could be measured in experiment.

A discovery of an electron electric dipole moment (e-EDM) would be unequivocal proof of new physics beyond the Standard Model. An EDM in an eigenstate of angular momentum is possible only if both Parity (P) and time reversal (T) are violated where T violation is, by the CPT theorem, the equivalent of CP violation. No EDM of any particle or system has yet been observed: all known CP violations (in the decays of the B and K$^0$ systems) are consistent with the Standard Model's Cabbibo-Kobayashi-Maskawa (CKM) mechanism. The CKM mechanism directly affects only the quark sector, and the CKM-generated e-EDM is extremely small.  It is estimated \cite{bern91,Pospelov-Ritz-2005} to be about $10^{-10} - 10^{-5}$ (depending upon assumptions about the number of neutrino generations and their masses) of the current e-EDM experimental limit of $2.6 \times 10^{-48}$ C-m ($1.6 \times 10^{-27}$ e-cm) \cite{regan02}. By improving the present e-EDM limit, constraints would be placed on many SMEs and possibly on current models of neutrino physics \cite{mohapatra05}.

Below we will discuss examples of space-based experiments testing for physics beyond the Standard Model.

\subsubsection{Probing the special theory of relativity in space-based clock-comparison experiments}

Searches for extensions of special relativity on a free-flying spacecraft or on the International Space Station (ISS) are known as ``clock-comparison'' experiments.\footnote{The clocks referred to here take several forms, including atomic and optical clocks, masers, and electromagnetic cavity oscillators. Some clocks may be cesium and rubidium atomic clocks, enhanced to exploit fully the low-gravity environment of space.  Each produces an exceptionally stable oscillating signal from energy-level transitions in alkali atoms.  Other ISS-based clocks may include masers generating stimulated microwave signals, and microwave cavities creating resonant radiation in small superconducting cavities.}   The basic idea is to operate two or more high-precision clocks simultaneously and to compare their rates correlated with orbit parameters such as velocity relative to the microwave background and position in a gravitational environment.  The SME allows for the possibility that comparisons of the signals from different clocks will yield very small differences that can be detected in experiment. 

Tests of special relativity and the SME were proposed by the Superconducting Microwave Oscillator (SUMO) group, those from the Primary Atomic Reference Clock in SPACE (PARCS) \cite{Heavner-etal-2001,Sullivan-etal-2005} and the Rubidium Atomic Clock Experiment (RACE) \cite{Fertig-etal-2005} originally slated for operation on the ISS in the 2005-07 time frame. SUMO, a cryogenic cavity experiment \cite{Lipa-etal-2005}, was to be linked with PARCS to provide differential red-shift and Kennedy-Thorndike measurements and improved local oscillator capability \cite{Sullivan-etal-2005}. Unfortunately, for programmatic reasons the development of these experiments was canceled by NASA in 2004.\footnote{\label{ref:factors}See the 2003 Report {\it ``Factors Affecting the Utilization of the International Space Station for Research in the Biological and Physical Sciences''} submitted by the NRC Space Studies Board's Task Group on Research on the ISS (The National Academies Press, 2003) and especially the list of the fundamental physics experiments to be flown on the ISS in the period of 2002-2008 on p. 75 of {\tt http://books.nap.edu/books/NI000492/html}} Currently, an experiment, called Atomic Clock Ensemble in Space (ACES), is aiming to do important tests of SME.  ACES is a European mission \cite{Salomon-etal-2001,Cacciapuoti-etal-2007} in fundamental physics that will operate atomic clocks in the microgravity environment of the ISS with fractional frequency stability and accuracy of a few parts in 10$^{16}$.  ACES is jointly funded by ESA and CNES and is being prepared for a flight to the ISS in 2013-14 \cite{Salomon-Cacciapuoti-Dimarcq-2007} for the planned mission duration of 18 months.

Optical clocks (see Section~\ref{sec:2.6}) offer improved possibility of testing the time variations of fundamental constants at a high accuracy level \cite{Marion-2003}.  Such measurements interestingly complement the tests of the local Lorentz invariance (LLI) \cite{Wolf-Petit-1997}  and of the universality of free fall to experimentally establish the validity of the EP. The universality of the gravitational red-shift can be tested at the same accuracy level by two optical clocks in free flight in a varying gravitational potential.  Constancy and isotropy of the speed of light can be tested by continuously comparing a space clock with a ground clock.  Optical clocks orbiting the Earth combined with a sufficiently accurate time and frequency transfer link, can improve present results by more than three orders of magnitude.

In general relativity, continuous space-time symmetries, such as the Lorentz symmetry, are part of gauge invariance, and as such, their violation can only be understood as the low-energy limit of some underlying spontaneous breaking \cite{Kostelecky-2004,Dvali-Pujolas-Redi-2007}. Any order parameter that spontaneously breaks Lorentz invariance will also couple to gravity ultimately distorting the geometry of space-time thereby severely constraining the value of that parameter.  A consistent Lorentz-breaking inevitably requires either going beyond the usual order parameters (which are different from ordinary cosmic fluids) or going beyond general relativity at large distances.  Such a connection suggests that there must be a strong correlation between detection of Lorentz violation and results of relevant cosmological observations. If no such correlation is found, then any discovery of Lorentz violation would indicate the existence of new physical sources or new gravitational dynamics at large distances.  Therefore, a consistent breaking of Lorentz-invariance is rather a profound effect that, if detected by an experiment, must be studied together with cosmological observations to determine its nature (see Section~\ref{sec:2.4} for additional details).

\subsubsection{Search for the Electron's Electric Dipole Moment}

Electron EDM experiments are a sensitive test for non-Standard Model sources of CP violation \cite{wolfenstein04}. New, non-CKM sources of CP violation, that directly affect leptons and that can give rise to a large e-EDM, are predicted by the SME and are within a reach of modern experiments. A non-CKM source of CP violation is thought to be necessary to generate the observed matter-antimatter asymmetry in the universe \cite{Sakharov-1967}. Within the range of future experiments, e-EDMs are predicted to arise from
couplings to new particles and with non-standard sources of CP violation. These
particles are, in some models, candidates for dark matter, or part of the
mechanism for generating the observed excess of matter over antimatter, or part
of the mechanism for generating neutrino mass. In fact, potentially observable
e-EDMs \cite{bern91, Pospelov-Ritz-2005, barr93} are predicted by supersymmetry
\cite{abel01}, multi-Higgs models, left-right symmetric models, lepton
flavor-changing models, technicolor models \cite{applequist04}, and TeV-scale
quantum gravity theories \cite{ADD-1998}. Split supersymmetry
\cite{arkani-hamed05, chang05, giudice06} predicts an e-EDM within a few orders
of magnitude of the present experimental limit and up to the present
experimental limit. Merely improving the present e-EDM limit would place
constraints on these Standard Model extensions and possibly on current models
of neutrino physics \cite{mohapatra05}.

Cold-atom-based experiments may be used to search for e-EDM.  As a first step a ground-based demonstration of a Cs fountain e-EDM experiment has been carried out at LBNL \cite{Amini-Munger-Gould-2007}.  Similar to an atomic clock, a cold-atom-based e-EDM experiment is more sensitive in the microgravity environment of space than on the ground.  Such an experiment may improve the current sensitivity to the e-EDM by several orders of magnitude.

Direct measurement of any SME effects would herald a new era, fundamentally changing the perspective on the fabric of special relativity.  The effect of such a discovery would include permanent changes in cosmology, high-energy physics, and other fields.

The current research efforts in this area are limited to ground-based laboratory work supported by the NSF. All previously available NASA funding to the space-based efforts described above was terminated after the 2004 cancellation of the ``Microgravity and Fundamental Physics'' program (see Appendix~\ref{sec:2app} for details); no NASA support for research in this area is currently available.

\subsection{\label{sec:2.6} Cold atom physics, new frequency standards and quantum technologies}

Studies of the physics of cold atoms and molecules have recently produced sensational scientific results and important inventions.  The field of Atomic, Molecular and Optical Physics has had an incredibly productive decade marked by Nobel Prizes awarded discoveries in laser cooling (1997), Bose-Einstein condensation and atom lasers (2001), laser-based precision spectroscopy and the optical frequency comb technique (2005).

Quantum principles are in the core of many advanced technologies used in high-accuracy experiments in fundamental physics.  One of the areas where the progress will certainly bring societal and technological benefits is the area of condensed quantum matter.  The new phenomena of interest is seen only at extremely low temperatures, or at very high densities, when particles such as electrons in some metals, helium atoms, or alkali atoms in atomic traps, form a pattern not in ``real'' space, but rather in momentum space.  This ``condensation'' into momentum space is known as Bose-Einstein condensation (BEC), which in strongly interacting systems, such as liquid helium, results in a complex phenomenon known as super-fluidity.  BEC in the ``paired'' electrons in some metals results in superconductivity, which has already revolutionized many technologies to date and is poised to produce many new applications in the future.

While gravitational and relativistic physics examine the most fundamental laws describing the universe on the large scale, it is equally important to look at the tiny building blocks of matter and how they manifest the same fundamental laws.  New techniques allow us to use laser light to cool and probe the properties of individual atoms as a starting point for exploration.  Working with individual atoms as test laboratories, researchers stand at the bridge between the smallest pieces of matter and the complex behavior of large systems.  

Furthermore, conducting these experiments in space allows one to remove the influence of gravity and manipulate matter freely, without having to counteract specimens ``falling'' within the instruments.  It is possible to study clouds of atoms, cooled by laser light to very near absolute zero yet freely floating without the forces that would be needed to contain them on Earth.  This unique realization of an atom nearly at rest in free space allows longer observation times and enables measurements with higher precision (see discussion of the experiments relying on cold-atom-based technologies in sections above).

Below we present several impressive examples of the new generation of quantum technologies and discuss their role for space-based experimental research in fundamental physics.

\subsubsection{\label{sec:clocks}Highly-accurate optical clocks}

Precision physics, in particular precision frequency measurements, has recently shown substantial progress with the introduction of new types of atomic clocks. Recent advances in accuracy that have turned the clocks into powerful tools in the development of applications in time and frequency metrology, universal time scales, global positioning and navigation, geodesy, gravimetry and others.

Present-day trapped-ion clocks are capable of a stability at the $10^{-15}$ level over a few hours of integration time \cite{Prestage-2007}.  Such performance allows simultaneous 1-way (down-) and 2-way (up- and down-) links between space and ground with a greatly reduced tropospheric noise contribution. These clocks are light-weight, have no lasers, cryogenics, or microwave cavities and are similar to a traveling-wave tube that is already on-board many spacecraft.  

Microwave atomic clocks with similar performance have also been space qualified \cite{Laurent-etal-2006} and are being prepared for a flight on the ISS \cite{Salomon-etal-2001,Cacciapuoti-etal-2007}. In fact, the current accuracy of fountain clocks is already at the $5 \times 10^{-16}$ level and expected to improve in the near future.

Optical clocks have already demonstrated fractional frequency stability of a few parts in $10^{17}$ at 1 second of integration time.  The performance of these clocks has strongly progressed in recent years, and accuracies below one part in $10^{17}$ are expected in the near future \cite{Oskay-etal-2006}.  Access to space provides conditions to further improve the clock performance potentially reaching the $10^{-18}-10^{-19}$ stability and accuracy level \cite{Rosenband-etal-2006,Rosenband-etal-2007}. In addition, the operation of optical clocks in space provides new scientific and technological opportunities \cite{Phillips-2007,Schiller-etal-2006} with far reaching societal benefits.

High-precision optical clocks can measure the gravitational red-shift with a relative frequency uncertainty of few parts in $10^{18}$, demonstrating a new efficient way of mapping the Earth gravity field at the cm level.  The universality of high accuracy time transfer is another important aspect of an optical clock. In particular, it is feasible that a few space-based high-precision clocks will provide a universal high precision time reference for users both ground-based and in space -- a highly accurate clock synchronization and time transfer that can not be achieved from the ground.  There will be a strong impact on various disciplines of Earth sciences that will greatly benefit from the use of clocks including studies of relativistic geodesy, Earth rotation, climate research, ocean research, earthquakes, tsunamis, and many others.  

There are several benefits of space deployment for the latest generation of atomic clocks.  Thus, it is known that thermal noise and vibration sensitivity are the two factors limiting the performance of the optical local oscillator \cite{Boyd-etal-2006,Ludlow-etal-2007}. By operating the clock in a quiescent space environment one can use longer cavity spacers to reach better short term stability for the local oscillator.  That will have a direct impact on spectral resolution as better spectral resolution will lead to smaller systematic errors.

\subsubsection{Femtosecond optical frequency combs in space}

The optical frequency comb (OFC) is a recently discovered method that enables comparison of the frequency stability of optical clocks at the level of $\delta f/f \sim10^{-19}$ \cite{Ma-etal-2004}. In the optical domain, using a frequency comb a group at NIST observed a beat note frequency stability at the $3\times 10^{-17}$ level between Al$^+$ and Hg$^+$ ions  \cite{Oskay-etal-2006}.  Where the stability of the microwave-to-optical link is concerned, with the help of a mode-locked femtosecond laser emitting a series of very short laser pulses with a well-defined repetition rate, an OFC can be generated that makes it possible to compare the microwave frequency of atomic clocks (about $10^{10}$~Hz) with optical frequencies (about $10^{15}$ Hz) with an accuracy of one part in $10^{15}$ \cite{Diddams-etal-2000,Fortier-etal-2007}. 

The OFC technique will lead to further improvements of Kennedy-Thorndike tests \cite{Kennedy-Thorndike-1932} and also tests of the universality of the gravitational red-shift \cite{Phillips-Ye-2007}.  In particular, the ability of OFC to  accurately measure differential frequency shifts across a wide electromagnetic spectrum would lead to improving results of these experiments. 
%
%
OFC has aided the development of better optical clocks.  With reduced vibration-related problems in the space environment, further improvements can be made in the traditional optical-cavity-based speed-of-light tests (see discussion in Section~\ref{sec:3}).  Instead of using two  orthogonally-oriented cavities to test the angle-dependent (Michelson-Morely) and angle-independent terms (Kennedy-Thorndike, velocity-dependent terms), using an optical clock one can do these tests with one cavity, while using the clock itself as the absolute reference. The fact that this would be an all-optical test is an additional benefit of the method. 

As an example of a possible experiment, one could use a rotating wide-bandwidth optical cavity.  As the cavity rotates in space, with its axis pointing to different gravitational sources, one can measure possible frequency shifts in blue, green, and red colors, or even all the way to microwave (group delay) frequencies, all in one setup.  These measurements could give an extra constraint with respect to the optical carrier frequency and improve the experimental sensitivity \cite{Ye-pc-2007}. There are other promising non-frequency comb proposals, for instance, those based on high finesse cavities \cite{Lipa-etal-2003,Stanwix-etal-2006,Wolf-etal-2004}.

OFCs enable a unique opportunity of simultaneous time-keeping and distance ranging.  Traditionally laser ranging involves measuring the delay of pulses reflected or transponded from the ranging target.\footnote{Lunar ranging from a reflector placed by lunar astronauts has, since the early 1970s, improved from several tens of centimeters to a level now of about a millimeter; see \cite{Williams-etal-2004,Williams-Turyshev-Murphy-2004,Murphy-etal-2007}.}  Use of the highly coherent light associated with modern femtosecond combs would allow  interferometry, supplemented by pulse delay ranging \cite{Ye-2004}.  The idea of absolute distance measurement within an optical fringe based on a comb can be easily implemented in space due to a significantly lower contribution from any dispersive medium there.  Substantial improvement in laser ranging would aid studies of relativistic gravity in the solar system as well as astrometry, geodesy, geophysics, and planetology \cite{Phillips-2007}.

\subsubsection{\label{sec:quantum-sensors}Atomic quantum sensors}

Atomic quantum sensors based on matter-wave interferometry, are capable of detecting very small accelerations and rotations (see discussion in \cite{Borde-1989,Peters-Chung-Chu-1999}).   The present day sensitivity of atom interferometers used as accelerometers  is $\delta a \sim 10^{-9}$~m/s$^2/\sqrt{\rm Hz}$ and as gyroscopes $\delta \omega \sim 6 \times 10^{-10}$~rad/s/$\sqrt{\rm Hz}$ \cite{Gustavson-Landragin-Kasevich-2000}.  These instruments reach their ultimate performance in space, where the long interaction times achievable in a freely-falling laboratory improve their sensitivity by at least two orders of magnitude (for review, \cite{Miffre-etal-2006}).  Possibilities of improvements in the measurement techniques with differential atom interferometry, including those beyond the standard quantum limit, are also investigated \cite{Eckert-etal-2006}.

Cold atom sensors in space may enable new classes of experiments such as testing gravitational inverse-square law at distances of a few microns, the universality of free fall and others.  Matter-wave interferometer techniques may lead to radically new tests of the EP using atoms as nearly perfect test masses \cite{Kasevich-Maleki-2003}, measurements of the relativistic frame-dragging precession  \cite{Jentsch-etal-2004},  the value of $G$ \cite{Bertoldi-etal-2006,Fixler-etal-2007} and other tests of general relativity \cite{Dimopoulos-etal-2007}.  Furthermore, cold atom quantum sensors have excellent sensitivity for absolute measurement of gravity, gravity gradients and magnetic fields as well as Earth rotation, and therefore find application in Earth sciences and in Earth-observing facilities \cite{Yu-etal-2006}.   Miniaturized cold-atom gyroscopes and accelerometers may lead to the development of autonomous navigation systems not relying on satellite tracking \cite{Coq-etal-2006,Canuel-etal-2006,Ketterle-etal-2007,Durfee-etal-2006}.

Today, a new generation of high performance quantum sensors (ultra-stable atomic clocks, accelerometers, gyroscopes, gravimeters, gravity gradiometers, etc.) is surpassing previous state-of-the-art instruments, demonstrating the high potential of these techniques based on the engineering and manipulation of atomic systems.  Atomic clocks and inertial quantum sensors represent a key technology for accurate frequency measurements and ultra-precise monitoring of accelerations and rotations \cite{Phillips-2007}.  New quantum devices based on ultra-cold atoms will enable fundamental physics experiments testing quantum physics, physics beyond the Standard Model of fundamental particles and interactions, special relativity, gravitation and general relativity \cite{Tino-etal-2007}. Because of the anticipated strong impact of these new devices on the entire area of precision measurements, the development of quantum technologies for space applications has also seen increased activity.\footnote{For instance, a recent European Workshop on ``Quantum mechanics for space,'' held at ONERA, Ch\^atillon, France, during  30 March - 1 April 2005 see {\tt http://qm-space.onera.fr/} and the related proceedings issue \cite{QM-2005}.}

In addition, studies on ultra-cold atoms, molecules and degenerate quantum gases (BEC, Fermi gases, and Bose-Fermi mixtures) are also steadily progressing \cite{Ketterle-2007,Ye-2007}.  BEC provides gases in the sub-nano-Kelvin range with extremely low velocities (i.e., at the micron per second level), that are ideally suited for experiments in a microgravity environment. Similarly, Fermi gases have no interactions at low temperatures which is very important for potential tests of the EP in space-based experiments. The atom chip technology  and compact interferometers now under development may provide a low power, low volume source for atomic quantum sensors \cite{Boyer-2007}.
 
In summary, it is clear that in recognition of such an impressive progress and also because of its uniquely strong potential for space applications, research in cold atoms and quantum technologies would greatly benefit from NASA (and, hopefully, multi-agency) support; the current  European effort in this area is an excellent example.\footnote{ESA has recently established  two programs to develop atomic clocks and atom interferometers for future missions in space, namely: ESA-AO-2004-100, ``Space Optical Clocks,'' coordinated by S. Schiller and ESA-AO-2004-064/082, ``Space Atom Interferometers,'' coordinated by G.\,M. Tino.} We emphasize that a coordinated multi-agency support could lead to significant progress in developing quantum technologies for space applications and will benefit the entire discipline of space-based research in fundamental physics.

\section{\label{sec:4} Discussion and recommendations}

The recent technological progress and the availability of the quiescent environment of space have placed fundamental physics in a unique position to address  some of the pivotal questions of modern science. The opportunity to gather important new knowledge in cosmology, astronomy, and fundamental physics stems from recent discoveries suggesting that the basic properties of the universe as a whole may be
intimately related to the physics at the very smallest scale that governs elementary particles such as quarks and other constituents of atoms.

The science investigations presented in this paper are focusing on the very important and challenging questions that physics and astronomy face today.  Space deployment is the common factor for these investigations; in fact, their science outcomes are more significant if the experiments are performed in space than on the ground. 

Because of the significant discovery potential offered by the space-based laboratory research in fundamental physics it would be beneficial to set aside some dedicated multi-agency funding to stimulate the relevant research and development efforts. Investments in this area of fundamental physics are likely to lead to major scientific advances and to the development of new 
technologies and applications that will strengthen national economic competitiveness and security.\footnote{Note that NASA was not included to receive a funding increase as a result of the American Competitiveness Initiative (ACI) (see {\tt http://www.ostp.gov/html/ACIBooklet.pdf}).  ACI provided additional authorization of \$160 million for basic science and research for FY 2007 to other agencies.}

Our specific recommendations are below:

\begin{center}
{\bf \small 1.~~~Include Fundamental Physics in the NAS' Decadal Survey in Astronomy and Astrophysics}
\end{center}
   
We recommend that the upcoming National Academy of Sciences' Decadal Survey in Astronomy and Astrophysics\footnote{In addition, the upcoming NAS' Decadal Survey in Physics should also have a strong responsibility for recommendations to NASA on space-based research in fundamental physics.} include space-based research in fundamental physics as one of its focus areas.  As new scientific discoveries and novel experimental approaches cut across the traditional disciplinary boundaries, the next survey should make physics an equal partner for strategic planning of future efforts in space.  The space-based research community in fundamental physics is ready to contribute to this important strategic planning process for the next decade, as evidenced by ``Quantum to Cosmos'' meetings.

\begin{center}
{\bf \small 2.~~~Establish an Interagency \\ Fundamental Physics Task Force}
\end{center}

For the near-term, we recommend that the Astronomy and Astrophysics Advisory Committee (AAAC)\footnote{The Astronomy and Astrophysics Advisory Committee (AAAC), see details {\tt http://www.ucolick.org/\~{}gdi/aaac/} and also NFS AAAC web-page {\tt http://www.nsf.gov/mps/ast/aaac.jsp}} establish an interagency ``Fundamental Physics Task Force (FPTF)'' to assess benefits of space-based efforts in the field, to identify the most important focus areas and objectives, and to suggest the best ways to organize the work of the agencies involved \cite{Marburger-2007}.  

The FPTF can help the agencies identify actions that will optimize a near- and intermediate-term synergistic fundamental physics programs carried out at NSF, NASA, DOE and other federal agencies\footnote{\label{agencies}The 2006 {\it ``Quantum to Cosmos''} (Q2C) workshop  (see meeting details at {\tt http://physics.jpl.nasa.gov/quantum-to-cosmos/}) demonstrated that, in addition to NASA, NSF, and DOE-SC, agencies like the Department of Commerce's National Institute of Standards and Technology (DOC/NIST) and others do benefit from laboratory physics research in space and, thus, could also sponsor the FPTF (see discussion in Ref.~\cite{Marburger-2007}).} and ensure progress in the development and implementation of a concerted effort towards improvement of our understanding of the fundamental laws of physics.  

Given the scientific promise of ``Quantum to Cosmos'' and also its inter-disciplinary nature, it is the right time for the National Science and Technology Council's Committee on Science\footnote{National Science and Technology Council (NSTC) at the Executive Office of the President of the United States, see website: {\tt http://www.ostp.gov/nstc/}} to include the new program under jurisdiction of the Interagency Working Group (IWG) on the ``Physics of the Universe'' and to ask the IWG to examine the investments required to support the new area of space-based laboratory research in fundamental physics and to develop priorities for further action.

\begin{center}
{\bf \small 3.~~~Establish a NASA-led program dedicated to space-based efforts in Fundamental Physics}
\end{center}

In the intermediate term, we recommend that NASA to establish a program dedicated to space-based efforts in fundamental physics and quantum technologies.  The new program, tentatively named  ``Quantum to Cosmos'' (Q2C), would complement NASA's {``Beyond Einstein''} program$^{\ref{be-program}}$ and could also  offer a ``space extension'' to NSF and DOE ground-based research efforts by providing unique opportunities for high accuracy investigations in fundamental physics. 

The program would focus on high-precision tests of relativistic gravity in space, searches for new physics beyond the Standard Model, direct detection and studies of gravitational waves, searches for dark matter, discovery research in astroparticle physics, and precision experiments in cosmology. It would also develop and utilize advanced technologies needed for space-based experiments in fundamental physics, such as laser transponders, drag-free technologies, atomic clocks, optical frequency combs and synthesizers, atom and matter-wave interferometers, and many others. 

The new program would allow NASA to fully utilize science potential of the ISS by performing carefully planned fundamental physics experiments on-board the space station -- which would be a direct response to the 2005 designation of the ISS as a national laboratory.\footnote{The 2005 NASA Authorization Act designated the U.S segment of the ISS as a National Laboratory and directed NASA to develop a plan to increase the utilization of the ISS by other federal entities, the research community and the private sector; see {\tt http://www.nasa.gov/mission\_pages/station/science/nlab/}. The majority of NASA's ISS research effort is focused on supporting the human exploration of space program, with only about 15\% going towards other activities. There is currently no NASA support for performing fundamental physics aboard the ISS. Thus, the designation of the ISS as a national laboratory rings untrue as long as major research areas, such as fundamental physics, are excluded from participating. The recent NASA Request for Information for Earth and Space Science Payloads on the ISS could be an important first step in the right direction.}

With its strong interdisciplinary research focus such a program would broadly cut across agencies, academia, and industry while also overlapping interests of several federal agencies, namely NASA, NSF, DOE/SC, DOC/NIST, NIH, and other. We believe that this program will allow the U.S. to re-gain a leadership position in fundamental physics worldwide.

\begin{center}
{\bf \small 4.~~~Enrich and Broaden NASA's Advisory Structure with Space-Based Laboratory Fundamental Physics}
\end{center}

We observe that ESA's FPAG$^{\ref{fpag}}$ is a good example of how to engage the fundamental physics community in space-related research activities, enrich and deepen the space enterprise, and also broaden ESA's advocacy base. NASA would benefit from access to a similar group of science advisors. The recently formed NASA Advisory Committee  (NAC)\footnote{\label{nac}The NASA Advisory Committee  (NAC), for details see webpage at {\tt http://www.hq.nasa.gov/office/oer/nac/}} does not have representation from the fundamental physics community, nor does the Astrophysics sub-committee\footnote{The webpage of the NAC's Astrophysics sub-committee: {\tt http://science.hq.nasa.gov/strategy/subcomm.html }}  of the NAC.  Therefore, we propose
\begin{itemize}
  \item To include members of the fundamental physics community in the NAC and/or NAC's Astrophysics sub-committee. Include fundamental physics in Astrophysics Division of the SMD and provide adequate representation in advisory structure. 
  \item  To consult with ESA on the possibility of appointing U.S. ex-officio members to ESA's FPAG.  Such participation could facilitate development of on-going (i.e., LPF, LISA) and future fundamental physics missions.

\end{itemize}

We believe that the recommendations presented above will benefit space-based research in fundamental physics and quantum technologies, the area of research with its unique science and strong technological potential.

\begin{acknowledgments}
We would like to express our gratitude to our many colleagues who have either collaborated with us on this manuscript or given us their wisdom.  We specifically thank Eric G. Adelberger, Peter L. Bender, Curt J. Cutler, Gia Dvali, John L. Hall, Wolfgang Ketterle, V. Alan Kosteleck\'y, Kenneth L. Nordtvedt, Douglas D. Osheroff, Craig Hogan, William D. Phillips, E. Sterl Phinney, Thomas A. Prince, Irwin I. Shapiro,  Jun Ye and Frank Wilczek who provided us with valuable comments, encouragement, support and stimulating discussions while this document was in preparation.

We are grateful to our European colleagues who benefited us with their insightful comments and suggestions regarding the manuscript. In particular, our gratitude goes to Orfeu Bertolami, Robert Bingham, Philippe Bouyer, Luigi Cacciapuoti, Thibault Damour, Hansjoerg Dittus, Ulrich Johann, Claus L\"ammerzahl, Ekkehard Peik, Serge Reynaud, Albrecht Ruediger, Christophe Salomon, Stephan Schiller, Mikhail Shaposhnikov, Guglielmo Tino, Andreas Wicht, and Peter Wolf.

We thank our colleagues at JPL for their encouragement, support, and advice regarding this manuscript.  We especially appreciate valuable contributions from Roger A. Lee, Moshe Pniel, Michael W. Werner and Jakob van Zyl of the Astronomy and Physics Directorate, Robert J. Cesarone of the Architecture and Strategic Planning Office of the Interplanetary Network Directorate and also from William M. Folkner and James G. Williams. Our gratitude also goes to Michael H. Salamon of NASA and Nicholas White of GSFC who have kindly provided us with many insightful comments and valuable suggestions on various aspects of the manuscript. 

The work described here, in part, was carried out at the Jet Propulsion Laboratory, California Institute of Technology, under a contract with the National Aeronautics and Space Administration.

\end{acknowledgments}

\appendix
\section{\label{sec:2}Advantages of carrying out physics research in space}

Historically, experiments in fundamental physics focused on laboratory efforts involving ground-based, underground and, more recently, balloon experiments.  Scientific progress in these experiments depends on clever experimental strategy and the use of advanced technologies needed to overcome the limits imposed by the environment typically present in Earth-based laboratories.  As a result, the use of sophisticated countermeasures needed to eliminate or reduce contributions from various noise sources increases the overall cost of the research.  Oftentimes the conditions in the Earth-based laboratories cannot be improved to the required levels of purity.

Although very expensive, access to space offers conditions that are not available on Earth, but are pivotal for many pioneering investigations exploring the limits of modern physics. We assert that, for many fundamental physics experiments, especially those aiming at exploring gravitation, cosmology, atomic physics, while achieving uttermost measurement precision, and increasingly so for high energy and particle astrophysics, a space-based location is the ultimate destination.

Considering fundamental physics experiments, our solar system is a unique laboratory that offers plenty of opportunities for discovery.  A carefully designed space-based experiment can take advantage of a number of factors that significantly improve its accuracy; some of these factors are listed below.

\subsection{Access to significant variations of gravitational potential and acceleration}

Some general relativistic effects (such as the gravitationally-induced frequency shift) vary with the potential.  Space enables far more precise tests because the change in potential can be much greater, and at the same time less affected by noise, than in any ground-based experiments.  Other effects (for example, those resulting from a violation of the EP or of local Lorentz and position invariances (see Sec.~\ref{sec:2.2} and~\ref{sec:2.1}) vary with the magnitude or direction of the acceleration.  These, and the range of frequencies at which they can be made to occur in the experiment frame, can greatly exceed the values accessible in corresponding ground-based tests.

The strongest gravity potential available in the solar system is that provided by the Sun itself.  Compared with terrestrial conditions, the Sun offers a factor of $\sim$3,000 increase in the strength of gravitational effects.  The corresponding gravitational acceleration near the Sun is nearly a factor of 30 larger than that available in ground-based laboratories.

Placing an experimental platform in heliocentric orbit provides access to conditions that are not available on the ground.  For instance, a highly eccentric solar orbit with apoapsis of 5 AU and periapsis of 10 solar radii offers more than 2 orders of magnitude in variation of the solar gravity potential and 4 orders of magnitude in variation in the corresponding gravitational acceleration, clearly not available otherwise.
Smaller benefits may be achieved at a lower cost in Earth orbit.   

In addition, ability to precisely track very long arcs of trajectories of test bodies in the solar system is another great advantage of space deployment.

\subsection{Greatly reduced contribution of non-gravitational sources of noise}

Compared with Earth-based laboratories, experiments in space can benefit from a range of conditions especially those of free-fall and also with significantly reduced contributions due to seismic, thermal and other sources of non-gravitational noise.

Microgravity environments ranging from $10^{-4} \, {\sl g}$ to $10^{-6} \, {\sl g}$ achieved with free-falling platforms enable new laser-cooling physics experiments and high accuracy tests of gravity with co-located clocks.  A long duration in a controlled free-fall environment and drag-free operations benefit many experiments.  It is expected that the next generation optical atomic clocks will reach their full potential of accuracy only in space (see Sec.~\ref{sec:clocks}).

Purely geodesic orbits that are achieved with drag-free spacecraft and for which effects of non-gravitational forces are reduced to $10^{-10} \, {\sl g}$ to $10^{-14} \, {\sl g}$ levels compared with terrestrial conditions are needed for precision tests of gravity and direct detection of gravitational waves.

\subsection{Access to large distances, velocities, and separations; availability of remote benchmarks and inertial references} 

Laser retroreflectors on the Moon, radio transponders on Mars, and radio-science experiments on-board remote spacecraft have vastly improved the accuracy of tests of relativistic gravity.  In the near future, optical- and atom-based quantum technologies could provide even higher accuracy for the next-generation of interplanetary laser ranging experiments.

Formation flying technologies, with spacecraft separated by distances from hundreds of meters to millions of kilometers, enable larger apertures, more complex focal plane assemblies, and longer interferometric baselines than are possible on the Earth. For example, a gravitational wave observatory with a baseline of several million km is possible only in space, opening up the window to study low-frequency gravitational waves.

Availability of an inertial reference frame oftentimes is one of the most critical requirements for precision tests of gravity.  Modern-day precision star-trackers and spacecraft attitude control systems allow the establishment of inertial reference and, thus, to carry out experiments in the inertial or quasi-inertial environment of the solar system.

\subsection{Access to vacuum conditions of space} 

Space deployment provides for a significant reduction of atmospheric interference with the propagation of optical, radio, and x-ray signals.   In fact, absence of air allows perfect optical ``seeing'' and avoids particle annihilation in antimatter searches.  Thus, space conditions allow for small point spread functions (PSF) (due to the absence of atmospheric blurring) and PSF stability (due to the thermally stable environment only available in space) needed for many precision measurements. As a result, the space vacuum allows the construction of instruments with unique architectures enabled by highly-accurate optical metrology, including image-forming instruments with large apertures and long-baseline interferometers.

\subsection{Availability of critical technologies}

Development of technologies needed for many fundamental physics experiments is a very challenging task; however, recent years have seen the maturation of a significant number of key technologies that were developed to take advantage of the unique conditions available in space.  Among them are high-precision accelerometers, drag-free control using He-proportional thrusters or small ion thrusters as actuators, ultra-stable lasers in space, He-dewars, cryo-coolers, superconducting detectors, high precision displacement sensors, magnetic spectrometers, small trapped-ion clocks, lightweight H-maser clocks and atomic clocks using laser-cooled atoms.  Many of these technologies have been space-qualified and some have already been flown in space, thereby paving the road for the development of many fundamental physics experiments.

Many of the advanced space technologies developed for gravitational experiments can be directly applied to other space sciences, including astrophysics, cosmology, astroparticle and atomic physics. Such technological cross-pollination allows other science disciplines to take advantages of space deployment opportunities, thereby stimulating the progress in many areas of space research.

\section{\label{sec:2app}Fundamental Physics in Space: Lessons from the Past and Prospects for the Future}

In the past, NASA had recognized the potential and importance of fundamental physics research conducted in space.  In particular, the program on ``Microgravity and Fundamental Physics'' that was established in 1996 was focused primarily on research to be conducted on the ISS, thereby, contributing to the science justification for the space station.\footnote{Among the most successful outcomes of this program that existed during 1996-2004 were the Lambda Point Experiment (LPE, 1992), Confined Helium eXperiment (CHeX, 1997), the Critical Fluid Light Scattering (ZENO, 1991, 1994) and Critical Viscosity Xenon (CVX, 1997) experiments. See details at {\tt http://funphysics.jpl.nasa.gov/} and {\tt http://funphysics.jpl.nasa.gov/technical/ltcmp/zeno.html}.}  After the publication of the {\it Roadmap for {\it Fundamental Physics in Space}} in 1999\footnote{An electronic version of the {\it Roadmap for  Fundamental Physics in Space} is avaiable at {\tt http://funphysics.jpl.nasa.gov/ technical/library/roadmap.html}}, the NAC$^{\ref{nac}}$ endorsed the Roadmap and recommended broader support dedicated to fundamental physics.

Following the initial success of the NASA's ``Microgravity and Fundamental Physics'' program and in recognition of the emergence of fundamental physics as a space discipline not served by other commissions, the Committee on Space Research (COSPAR), established Commission H: Fundamental Physics in Space in 1996.\footnote{Please see COSPAR's webpage, {\tt http://www.cosparhq.org/}, for information on Commission H on Fundamental Physics in Space.}  Since the time of its inception, the contributions from Commission H at the biannual COSPAR meetings has been steadily growing.

As for NASA efforts in the field, a suite of missions and space-based experiments representing many areas of physics were developed and planned for flight in the period of 2002-2008.$^{\ref{ref:factors}}$ However, in 2004, following the NASA reorganization in a response to the {\it U.S. New Space Exploration Initiative}, Code U became part of the new NASA Exploration Systems Mission Directorate (ESMD). With no clear alignment to the ESMD's objectives, the Code U's fundamental physics budget was decimated within four months and the entire Program was terminated in September 2006.

In contrast, the ESA's science program, through persistent efforts by the fundamental physics community \cite{Huber-2007} has an established ``Fundamental Physics Program'' highlighted by development of several missions which are nearing flight-ready status.\footnote{Notably, the MicroSCOPE and ACES missions.} ESA's recently initiated Cosmic Vision 2015-2025 program seeks to respond to the most important and exciting scientific questions that European scientists want to address by space missions in the time-frame 2015-2025.$^{\ref{esa-cv}}$ In fact, Cosmic Vision marks a significant breakthrough for fundamental physics: for the first time, a major space agency has given full emphasis in its forward planning to missions dedicated to exploring and advancing the limits of our understanding of many fundamental physics issues, including gravitation, unified theories, and quantum theory \cite{Schutz-2005}.\footnote{In October 2007 several candidate missions have been selected for a consideration for launch in 2017/2018, see {\tt http://sci.esa.int/science-e/www/object/index.cfm?fobject id=41438}. Although, no fundamental physics mission was chosen at this time, the Cosmic Vision opportunity strongly motivated the work in the entire area of space-based experimental research.}  NASA would benefit from a similar bold and visionary approach. Proposals submitted to the call for ``Astrophysics Strategic Mission Concept Studies'' of ROSES\footnote{NASA Research Opportunities in Space and Earth Sciences (ROSES-2007), see details at {\tt http://nspires.nasaprs.com/}.} could lead to such an opportunity and potentially result in a dedicated space-based laboratory experiment in fundamental physics. 

\subsection{\label{sec:app2-probes}Dedicated missions to test relativistic gravity}

There are two distinct approaches to organize research in fundamental physics in space -- either a dedicated space mission to perform an experiment or a smaller scale experiment as a part of a planetary exploration mission.  Below we discuss these approaches in some detail.

NASA has successfully launched two PI-led missions devoted to experimental tests of general relativity, namely Gravity Probe A\footnote{Please see {\tt http://en.wikipedia.org/wiki/Gravity\_Probe\_A}} and Gravity Probe B.

The first convincing measurement of the third of Einstein's proposed tests of general theory of relativity -- the gravitational red-shift -- was made in 1960 by Pound and Rebka. The Pound-Rebka experiment was based on M\"ossbauer Effect measurements between sources and detectors spanning the 22.5 m tower in the Jefferson Physical Laboratory at Harvard.  In 1976, Gravity Probe A exploited the much higher ``tower'' enabled by space; a sub-orbital Scout rocket carried a Hydrogen maser to an altitude of 10,273 km and a novel telemetry scheme allowed comparison with Hydrogen masers on the ground.  The clocks confirmed Einstein's prediction to 70 ppm. More than 30 years later, this remains the most precise measurement of the gravitational red-shift \cite{Will-2006}.

Gravity Probe B was launched on April 20, 2004. The goal of this experiment is to test two predictions of general relativity by means of measuring the spin direction of gyroscopes in orbit about the earth.  For the 642 km high GP-B orbit, GR predicts two gyroscope precessions, the geodetic effect with a rate of 6.606 arcsec per year and the frame dragging effect with a rate of 39 milliarcsec per year.  A polar orbit is chosen so the two effects would occur at right angles and could be independently resolved.   The science instrument was housed in the largest helium dewar ever flown in space.  The helium lifetime set the experiment duration; the orbital setup and science data phase lasted 17.3 months, exceeding requirements.

GP-B was the first spacecraft with 6 degree of freedom active control: translation, attitude, and roll.  He boil-off gas proportional thrusters provided control actuation enabling the ``drag free'' control system to reduce cross tract acceleration to $10^{-11} \, {\sl g}$. The requirements on charge control, magnetic shielding, and pressure were all met with margin. Flight data confirm that the GP-B gyroscope disturbance drift rates were more than a factor of a million times smaller than the best modeled navigational gyroscopes. The small size of the relativity effects under test levied extreme requirements on the mission; their successful demonstration in space has yielded and will yield many benefits to future fundamental physics missions in space. Final results are pending the completion of the data analysis scheduled for early 2008.

\subsection{\label{sec:app2-mo}Missions of Opportunity on planetary missions}

In addition to the development of dedicated missions, another practical way to conduct fundamental physics experiments is to fly advanced instruments as Missions of Opportunity (MO) on planetary missions and the ISS.

The LLR experiment that was initiated in 1969 by the Apollo 11 astronauts placing laser retroreflectors on the lunar surface \cite{Williams-etal-2005} is an excellent example of a successful MO.  The resulting fundamental physics experiment is still active today and is the longest running experiment in the history of space science 
(see Sec.~\ref{sec:var-fun-const}).

One can further improve LLR-enabled science by delivering to the moon either new sets of laser retroreflector arrays or laser transponders pointed at Earth or both types of these instruments \cite{Turyshev-Williams-2007}.  A geographic distribution of new instruments on the lunar surface wider than the current distribution would be a great benefit; the accuracy of the lunar science parameters would increase several times.  A bright transponder source on the moon would open LLR to dozens of satellite laser ranging stations which cannot detect the current weak signals from the moon. This would greatly benefit LLR -- the living legacy of the Apollo program -- and  would also enhance the science outcome of the new lunar exploration efforts. 

Highly-accurate measurements of the round-trip travel times of laser pulses between an observatory on the Earth and an optical transponder on Mars could lead to major advances in gravitation and cosmology, while also enhancing our knowledge of the Martian interior.  Technology is available to conduct such measurements with picosecond-level timing precision, which could translate into mm-class accuracies achieved in ranging between the Earth and Mars.  Similar to its lunar predecessor, the resulting Mars Laser Ranging experiment could become an excellent facility to advance fundamental physics.

Other examples of successful MOs are the recently conducted gravity experiment on the  Cassini mission performed on its way to Jupiter during one of the solar conjunctions \cite{Bertotti-Iess-Tortora-2003} and a similar experiment that is planned for the ESA's {BepiColombo} mission to Mercury \cite{Iess-Asmar-2007}.

In general, research in fundamental and gravitational physics will greatly benefit from an established mechanism to participate on planetary missions as MOs that will offer sorely needed space deployment opportunities.


\end{document}